\documentclass[9pt,journal]{IEEEtran}

\usepackage{amsmath,epsfig}
\usepackage{amsfonts,amssymb}
\usepackage{fixmath}
\usepackage{dsfont}
\DeclareMathOperator*{\argmin}{arg\,min}

\usepackage{color}

\ifCLASSINFOpdf

\else
''''
\fi

\begin{document}

\title{Navigation domain representation for interactive multiview imaging}

\author{Thomas~Maugey,~\IEEEmembership{Member,~IEEE,}
Ismael~Daribo,~\IEEEmembership{Member,~IEEE,}
Gene~Cheung,~\IEEEmembership{Senior Member,~IEEE,}
        and~Pascal~Frossard,~\IEEEmembership{Senior Member~IEEE}
\thanks{
This work has been partly supported by the Swiss National Science Foundation, under grant 200021-126894.}
\thanks{T. Maugey and P. Frossard are with Ecole Polytechnique F\'ed\'erale de Lausanne
(EPFL), Signal Processing Laboratory - LTS4, CH-1015 Lausanne, Switzerland. Email: thomas.maugey@epfl.ch, pascal.frossard@epfl.ch}
\thanks{I. Daribo(*) and G. Cheung are  with National Institute of Informatics, 2-1-2, Hitotsubashi, Chiyoda-ku, Tokyo, Japan 101-8430. E-mail: daribo@nii.ac.jp, cheung@nii.ac.jp}
\thanks{(*) now at the European Patent Office}
}

\maketitle

\begin{abstract}
Enabling users to 
interactively navigate through different viewpoints of a static scene is a new interesting functionality in 3D streaming
systems. While it opens exciting perspectives towards rich multimedia
applications, it requires the 
design of 
novel representations and
coding techniques in order to solve the new challenges imposed by interactive
navigation. In particular, the encoder must prepare \emph{a priori} a compressed media stream
that is flexible enough to enable the free selection of multiview navigation
paths by different streaming media clients. Interactivity clearly brings new design
constraints: the encoder is unaware of the exact decoding process, while the decoder has to reconstruct information from incomplete subsets
of data since the server can generally not transmit images for all possible
viewpoints due to resource constrains. 
In this paper, we propose a novel multiview data representation that
permits to satisfy bandwidth and storage constraints in an interactive multiview streaming system. In particular, we 
partition the multiview navigation domain into segments, 
each of which is 
described 
by a reference image (color and depth data) and some 
auxiliary information. The auxiliary information enables the client 
to recreate any viewpoint in the navigation segment 
via view
synthesis. The decoder is then able to navigate 
freely in the segment
without further data request to the server; it requests additional data only
when it moves to a different 
segment. We discuss the benefits of this novel
representation in interactive navigation systems and further propose a method to optimize the partitioning of
the navigation domain into independent segments, under bandwidth and storage
constraints. Experimental results confirm the potential of the proposed
representation; 
namely, our system leads to similar compression
performance as classical inter-view coding, while it provides the high level
of flexibility that is required for interactive streaming. Due to these unique properties, our new framework
represents a promising solution for 3D data representation in
novel interactive multimedia services.
\end{abstract}


\section{Introduction}

\begin{figure}[htb]
  \centering
 \centerline{\epsfig{figure=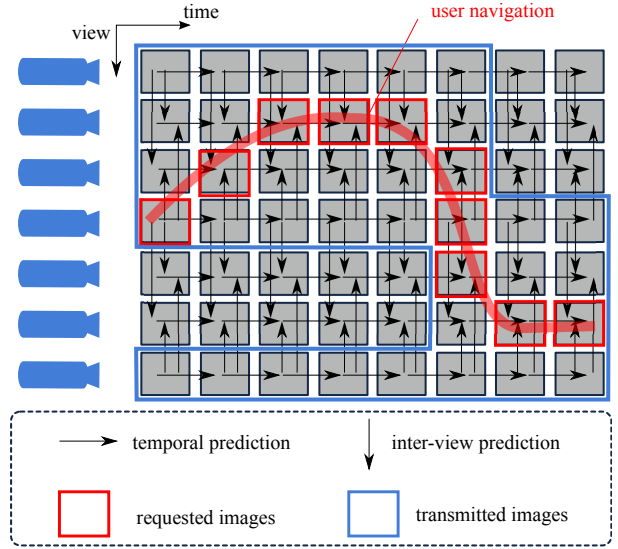,width=8cm}}
\caption{Traditional multiview prediction structures are not adapted for interactive navigation: for a given user navigation, more frames than requested are needed at the receiver side because of the heavy prediction structure.}
\label{fig:navigation}
\end{figure}

In novel multimedia applications, three dimensional data
information can be used to provide interactivity to the receiver,
and users can freely change viewpoints on their 2D displays. It enables the
viewer to freely adapt his viewpoint to the scene content and provides a 3D
sensation during the view navigation due to the look around effect
\cite{Muller_K_2011_pieee_tdv_rudm,Smolic_A_2005_pieee_int_tdvrct}. The design
of such an interactive system necessitates however the development of new techniques
in the different blocks of the 3D processing pipeline, namely acquisition
\cite{Benzie_P_2007_tcsvt_sur_tdtvtt}, representation
\cite{Alatan_A_2007_tcsvt_sce_rttdtvs}, coding
\cite{Smolic_A_2007_tcsvt_cod_atdtvs}, transmission
 \cite{Vetro_A_2011_tb_tdt_cst} and rendering
\cite{Holliman_N_2011_tb_thr_ddraa}. Solutions that are classically used for
multiview video transmission \cite{Wiegand_T_2003_tcsvt_ove_hvcs} are no
longer effective since they consider the transmission of an entire set of
views, which is not  ideal for interactive systems with delay and bandwidth
constraints. Fig~\ref{fig:navigation} illustrates that traditional compression methods introduce too many inter-frame dependencies while 
interactive systems should ideally transmit the requested views
only. Hence, the challenge is to build a representation that exploits the correlation
between multiview images for effective coding, but that is able at the same time to  satisfy the different users' navigation requests
without precise knowledge of the actual data available at decoder. With the classical
compression techniques based on inter-image prediction with motion/disparity
estimation, the problem can be solved with naive approaches in two specific scenarios. Firstly, if
the server is able to store all the  possible encoding prediction paths
in the multiview data, the user can receive only the required frames
(with a prediction path corresponding to its actual navigation) at low
bitrates. Secondly, it is also possible to implement a real-time encoding
(and thus real-time inter-image prediction) \cite{Lou_JG_2005_real_timvvs}
depending on the actual user position. However these two solutions do not scale with
the number of users and are therefore not realistic in practical settings. The
challenge for realizing an interactive multiview system for the streaming of static 3D scenes is thus twofold: i) to decrease the storage size without penalizing
too much the transmission rate and ii) to encode data \emph{a priori}
with random accessibility capabilities, thus avoiding computation cost associated with
a real-time encoder for each client. This further has to be done while considering of the complete system, from
the data capture to the view rendering blocks, including representation and
coding strategies. 

\begin{figure}[htb]
  \centering
 \centerline{\epsfig{figure= 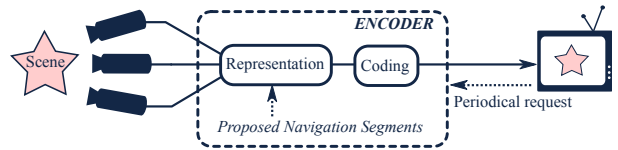,width=8cm}}
\caption{The encoder in a multiview system is composed of two blocks: data representation and coding. The  navigation segments proposed in this paper offers an alternative data representation method to classical multiview image representations.}
\label{fig:RepresentationAndCoding}
\end{figure}

 In this work, we build on \cite{Maugey_T_2012_picip_con_vsimi} and propose a radically new design for interactive navigation systems, which is supported by a flexible data representation method for static 3D scenes. 
The proposed solution achieves a high quality free-viewpoint navigation experience by limiting the data redundancies in the representation itself. An encoder typically has two means of reducing the data redundancy, as depicted in Fig.~\ref{fig:RepresentationAndCoding}. Traditional methods adopt a multiview representation and decrease the data size by improving the coding techniques \cite{Ohm_jr_jct3v_wp3dsd,Suzuki_T_2013_jvt3v_eidmdt,Rusanovskyy_D_2013_jct3v_tdavctm}. Although these methods are efficient from a compression perspective, they are not suitable for interactive scenarios since they introduce too many dependencies between the viewpoints, as illustrated in Fig.~\ref{fig:navigation}. In this work we rather focus on designing a novel representation framework, while previous works  rely on coding to reduce redundancy. Our novel data representation framework facilitates interactive navigation at decoder by providing data random access, without sacrificing much on coding efficiency.
Instead of optimizing data representation for a small set of predefined
viewpoints, we rather consider that free viewpoint navigation is described by
a navigation domain that contains all the possible virtual camera locations. The
navigation domain (ND) is divided into sub-domains called \emph{navigation
  segments}, which are transmitted to the decoder upon request. Upon reception
of data of a navigation segment, the decoder can independently create any
virtual view in this sub-domain without further request to the server. This
provides  flexible navigation capabilities to the receiver. But it also implies a
complete change in the data representation in order to limit storage and
bandwidth costs. Each navigation segment is thus represented with a reference
frame and some auxiliary information. The auxiliary information carries, in a
compact form, the innovation inherent to  new viewpoints and permits to
synthesize any view in the navigation segment with help of the reference frame.  We further propose to optimize the partitioning of the navigation
domain under rate and storage constraints. 
We finally illustrate the
performance of our system on several datasets in different
configurations. We observe that the proposed data representation achieves good
streaming performance that competes with  MVC-type inter-view prediction approaches, but at the same time offers high flexibility for interactive user
navigation. This new method provides a promising solution for the
design of 3D systems with new modes of interactions and rich quality of
experience.

A few solutions in the literature try to optimize the trade-off between
storage, bandwidth and interactive experience in multiview systems. A first category of methods
optimize  switching between captured views only. In other words, they adapt
the structure of the inter-view predictions in order to provide interactivity
at a moderate cost. Some of these methods are  inspired by the techniques that
have been developed to provide interactivity for monoview video. For example
the concept of SP/SI frames  \cite{Karczewicz_M_2003_tcsvt_sp_sifdh} is
adapted in \cite{Chen_Y_2009_jadvsp_eme_mvcstdvs} for view-switching. Other
works propose to modify the prediction structure between the frames
\cite{kurutepe_E_2007_tcsvt_cli_dssmvitdtv,Tekalp_AM_2007_ieee-spm_tdt_oipesmv}
by predicting the user position with the help of Kalman filtering. The authors
in \cite{Liu_Y_2010_jvci_rdo_ismvme} propose to store multiple encodings on
the server and to adapt the transmission to the user position. This is however
very costly in terms of storage. In
\cite{Kimata_H_2004_ntt_fre_vvcumvc,Shimizu_S_2007_tcsvt_vie_smvcutdwdm}, the
multi-view sequence is encoded with a GoGOP structure, which corresponds to a
set of GOPs (Group of Pictures). The limitation of such methods is a fixed
encoding structure that cannot be easily adapted to different
system configurations and application scenarios. In \cite{Cheung_G_2009_ipvw_gen_rfsims}, the problem is
formulated so that the proposed view-prediction structure reaches an 
optimal trade-off between storage and expected streaming rate. The possible 
types of frames are intra frames and predicted frames (with the storage of 
different motion vectors and residuals).  Some other techniques
\cite{Cheung_G_2011_tip_int_ssmvurfs,Petrazzuoli_G_2011_picip_usi_dscdibriimva,Xiu_X_2011_picip_fra_soimvsbnd}
rely on the idea of combining distributed source coding and inter-view
prediction for effective multiview switching. They propose an extension of the
view switching methods in a monoview framework
\cite{Cheung_NM_2006_pvicip_vid_cfpobdsc}. 
Unfortunately, all of these solutions remain limited since they restrict the
navigation to a small subset of views (the captured ones, generally not
numerous), which results in abrupt, unnatural view-switching
experience. Moreover, they cannot directly be extended to a system that
provides smooth navigation  through the whole scene (with a higher
number of achievable viewpoints).

A second category of methods try  to offer free viewpoint navigation by
considering a higher number of achievable views at the receiver. It could be
obtained by simply increasing the number of captured views, which is not
feasible in practice and not efficient in terms of redundancy in the
representation. Some solutions \cite{Xiu_X_2012_tmm_del_cimvfvs} extend the
previously mentioned techniques  by introducing  virtual view synthesis at the
decoder. However, they remain inefficient since the obtained virtual view
quality is low and the user navigation capacity is still limited. 
Contrary to what is assumed in these methods, virtual view synthesis algorithms do not only require two reference viewpoints (color+depth). Some occlusions may remain and need to be filled by inpainting algorithm. We claim in this paper that these techniques are limited as long as no further information is sent. In that sense, interactivity problem cannot be solved by simply sending reference viewpoints at strategical places.
Other
methods introduce high redundancy in the scene description by using a light
field representation
\cite{Tanimoto_M_2011_ieee-spm_fre_vtv,Tanimoto_2012_ieee-spm_ftv_fvt}. They
sample the navigation domain very finely, concatenate all the images and
finally model the light rays of the scene. The view rendering performed at the
receiver side with such light fields has a better quality and enables quite
a smooth navigation. However, the navigation path is pretty constrained and the data
representation does not achieve good compression performance. In general, all
the solutions that offer a large number of possible views have \emph{inherent
  redundancies in the representation}, which results in an inefficient
streaming system. It is finally important to note that none of the methods in the literature use an end-to-end system design approach. For example, while optimizing the coding techniques, almost none of the above works consider the constraints of the data rendering step.  It results in data blocks with strong dependencies, which are unfortunately not optimized for interactive navigation. 
The advantages of the novel framework proposed in this paper, as compared to the two 
aforementioned categories of previous approaches, are: i) unlike category 1,
it enables synthesis of a high number of virtual views, thus enriching
the interactive navigation experience, and ii) unlike category 2, it drastically
reduces the representation redundancy in the coded data, which leads to
lower streaming rate per navigation path (as compared to MVC-type inter-view prediction approaches), \textit{without} introducing inter-dependency in the compressed
media, that would reduce the flexibility for interactive data access.

The paper is organized as follows. In Sec.~\ref{sec:framework}, we introduce our novel framework for interactive multiview navigation. Then, we expose in Sec.~\ref{sec:partitioning} our solution to optimize the partitioning of the navigation domain. Finally, in Sec.~\ref{sec:exp}, we present different simulations results that validate the proposed approach.

\section{Interactive multiview navigation}\label{sec:framework}

\subsection{System overview}\label{sec:overview}
\begin{figure}[htb]
  \centering
 \centerline{\epsfig{figure= 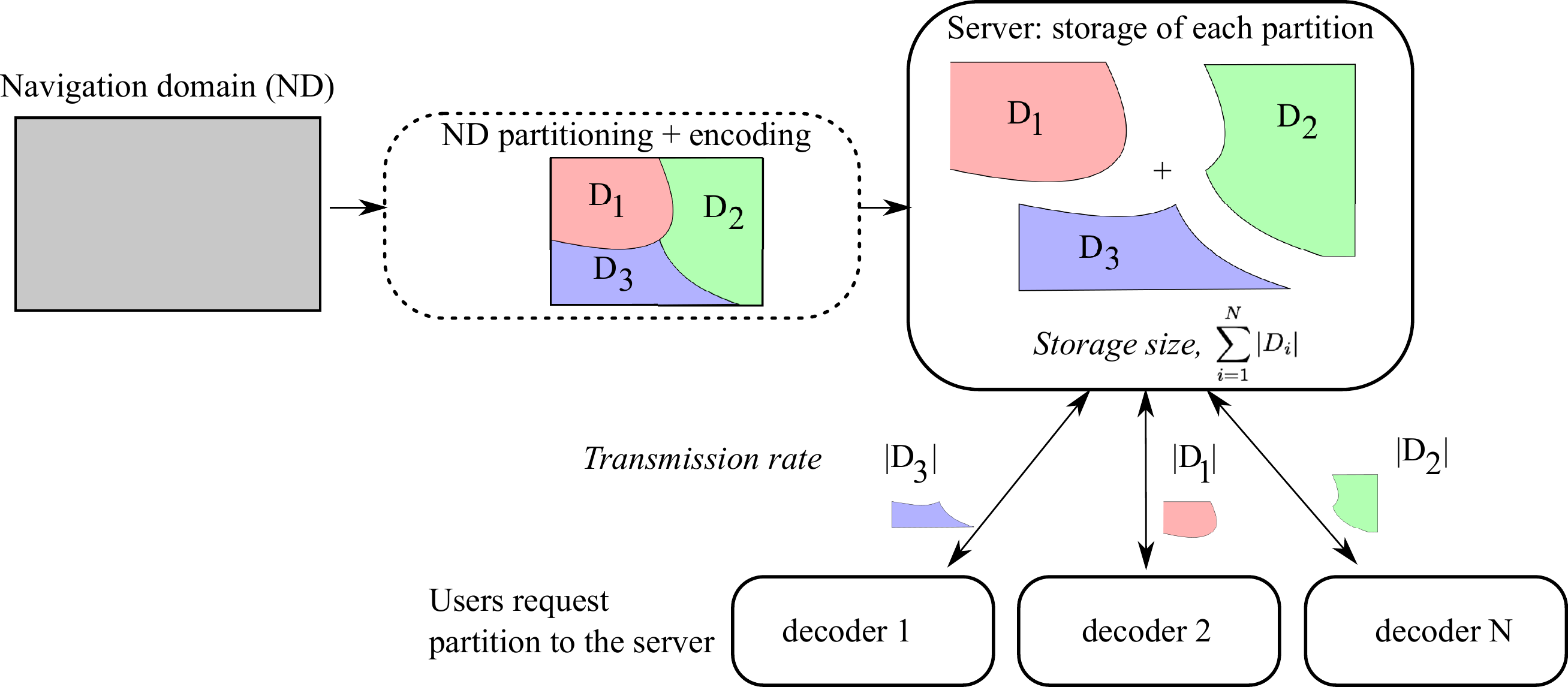,width=8cm}}
\caption{The navigation domain is partitioned into navigation segments, and each navigation segment is encoded and stored on a server. Users interact with the server to request the navigation segments needed for the navigation.}
\label{fig:System}
\end{figure}

In an interactive system, the user is able to freely navigate among a large set of viewpoints, in order to observe a static scene from different virtual camera positions. It generally means that the user has to communicate with a server and  request data that permits reconstruction and rendering of the desired virtual views on a 2D display. Let us consider a navigation domain constituted by a set of viewpoints. Our system relies on a novel data representation method that goes beyond the common image-based representation and rather considers  the global navigation domain as a union of different navigation segments. In more details, let us consider that the navigation domain is divided into $N_V$ navigation segments, which are each coded in a single data $D_i$, in the form of  one reference image and some auxiliary information (see Fig.~\ref{fig:System}). We further consider that a server stores all the $D_i$'s, with a  storage cost of $\sum_{i=1}^{N_V} |D_i|$, where $|D_i|$ is the size in bits of the data $D_i$. A user who navigates among the viewpoints regularly transmits its position to the server. If a user in a navigation segment $i$ comes close to the border of another navigation segment $j$, the server transmits the data $D_j$ to the user, which increments the reception rate cost by $|D_j|$. We see that, if the number of partitions $N_V$ in the navigation domain is large, the segment size $|D_i|$  decreases, but the number of user requests to the server increases if the navigation path is unchanged. On the contrary, if $N_V$ is low, the number of requests to the server decreases, but the user has to receive large segments $D_i$. We clearly see that $N_V$ should be determined carefully, taking into account both the bandwidth and storage constraints.

We notice that the communication between server and user is quite simple in our system. It has to deal with data transmission only when the user's navigation path gets  close to the borders of the navigation segments. Hence, our system scales pretty well with the number of users. Moreover, if the number of users becomes very high, one can consider a system with multiple replicas of the server.

\subsection{Navigation with 2D images}\label{sec:framework1}
\begin{figure}[htb]
  \centering
 \centerline{\epsfig{figure= 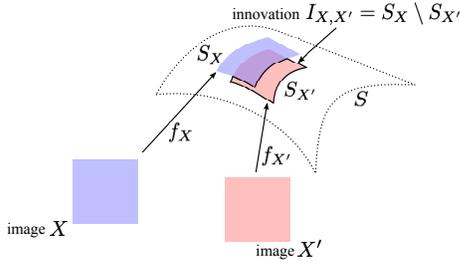,width=6cm}}
\vspace{0.5cm}
\caption{Illustration of visibility of scene elements in the images $X$ and $X'$. The innovation of $X'$ with respect to $X$ is represented with the black boundary.}
\label{fig:neighborhood}
\end{figure}

We provide now a formal description of the interactive multiview framework that we propose in this paper. We consider a system that captures and transmits data of a static 3D \textit{scene} $S$ to clients that can reconstruct 2D images of the scene for navigation, \emph{i.e.}, view-switching along certain directions. The scene $S$ is described as a countable set of random variables $s_i$ taking  values in $C^3$, where $C$ is the set of possible color values  (\emph{e.g.}, $C^3$ is $[0, 255]^{3}$)\footnote{In this work we make the Lambertian hypothesis, \emph{i.e.}, we assume that a voxel reflects the same color even when viewed from different viewpoints.}. Each of these random variables can be seen as a voxel in the 3D space \cite{CohenOr_D_2002_jgmip_fun_sv}. The decoder reconstructs observations of the scene at different viewpoints. These observations are 2D \emph{images} that correspond to finite sets of $N$ random variables $x_i$ taking their values in $C^{3}$. The observation of the 3D scene from one particular viewpoint gives an image $X$ that is obtained with a  \textit{projection function} associated to $X$. Since the depth information is known, we  define the back projection function that associates a pixel of an image to a 3D point in the scene:
\begin{align*}
f_X : X& \rightarrow  S \\
	   x& \rightarrow  s = f_X(x).
\end{align*}
This projection function depends on the distance between objects and the camera plane (\emph{i.e.}, depth) and on the extrinsic and intrinsic parameters of the camera \cite{Muller_K_2011_pieee_tdv_rudm,Tian_D_2009_pspie_vie_sttdv, Muller_K_2008_jivp_vie_satdvs,website_VVS}. 
In this work, we assume that each pixel in $X$ maps to a single voxel in 3D space $S$, and reciprocally, each voxel in $S$ maps to at most one pixel in $X$ (in other words, $f_X$ is a bijection of $X$ in $f_X(X)\subset S$). This assumption is correct as long as the 3D scene is sampled at a sufficiently high resolution, which is the scenario that we consider in the following. 
Not all the elements of $S$ can be seen from one viewpoint. We call $S_X = f_X(X)$ the finite subset of $S$ whose elements are mapped to elements of $X$. This is the set of elements of S that are \textit{visible} in $X$.  It naturally depends on the viewpoint. Our objective is to deliver enough information to the decoder, such that it can reconstruct different images of the scene. At the same time, the images from different viewpoints have a lot of redundancy. Ideally, to reconstruct an image $X'$ knowing the image $X$, it is sufficient for the decoder to receive the complementary, non-redundant information that is present in $X'$ (but not in $X$). We define it as the \emph{innovation} of $X$ with respect to $X'$: $I_{X,X'} = S_X \setminus S_{X'}$ (see Fig.~\ref{fig:neighborhood}).  
This innovation is due to two classical causes in view switching. First, \emph{disocclusions} represent the most complex source of innovation. They are due to pixels that are hidden by a foreground object or that are out of the camera range in the first view and become visible in the second view. The disocclusions are generally not considered at the encoder  in the literature. Existing schemes consider that they can be approximated by inpainting \cite{Criminisi_A_2004_tip_reg_forebii,Daribo_I_2010_mmsp_dep_aiinvs} or partially recovered via projection from other views \cite{Tian_D_2009_pspie_vie_sttdv}. Although the performance of inpainting techniques is improving, there still exists a problem with new objects or with frame consistency (especially when  neighboring frames are not available in interactive systems). This problem should be handled at the encoder and data to resolve disocclusions should also be sent to the decoder. We propose below a new data representation method that addresses this problem.

Second, innovation can also be generated by some new elements that appear due to a change in object resolution, \emph{i.e.}, when an object is growing from one viewpoint to another one. In other words, two consecutive pixels representing the same object in $X$ could map to two non-consecutive ones in $X'$ (even if they still describe the same object), leaving the intermediate pixels empty with no corresponding pixels in the reference view. This is due to the bijection assumption introduced above. However, we have chosen to restrict our study to the handling of disocclusions and we assume that these missing pixels due to resolution changes are recovered by a simple interpolation of the neighboring available pixels (of the same object). This assumption remains reasonable if we consider a navigation without large forward camera displacements. This is actually what is classically considered in view synthesis studies \cite{Tian_D_2009_pspie_vie_sttdv, Muller_K_2008_jivp_vie_satdvs}. Therefore, in the experiments, we will only consider navigation trajectories that remain at a similar distance from the scene.

\subsection{Navigation domains} \label{sec:ND}
\begin{figure}[htb]
  \centering
\vspace{0.5cm}
 \centerline{\epsfig{figure= 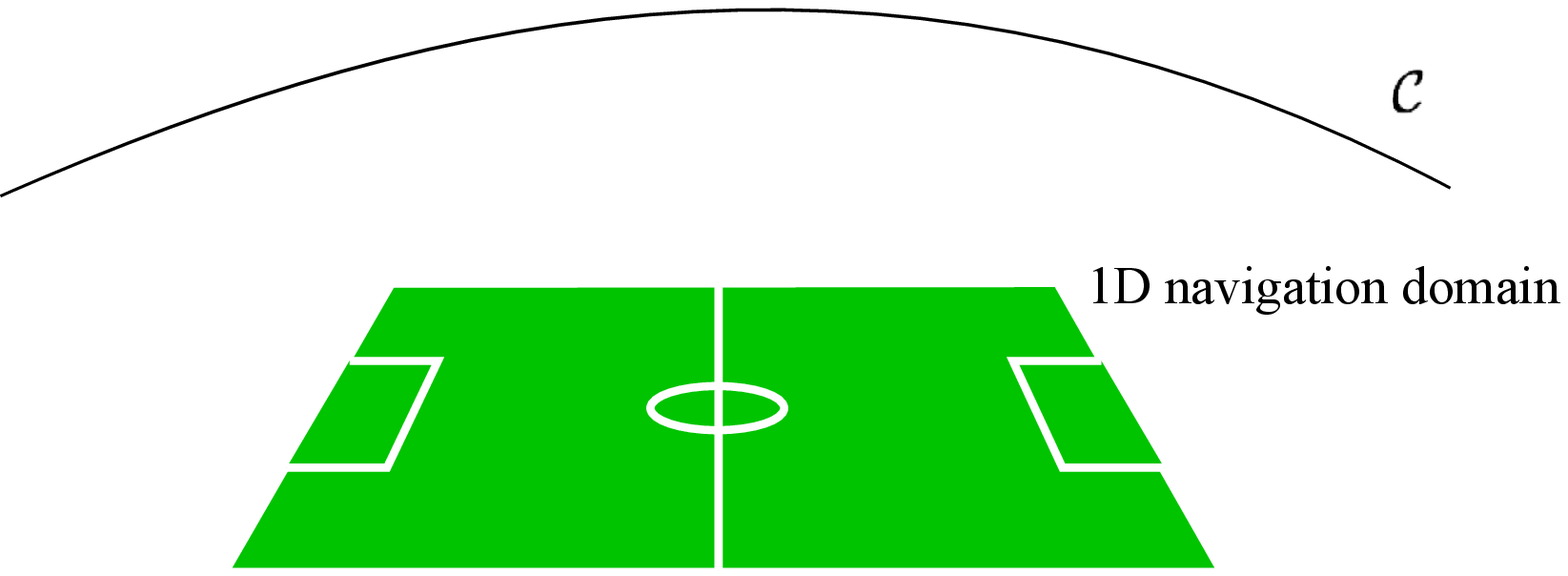,width=6cm}}
 \centerline{\epsfig{figure= 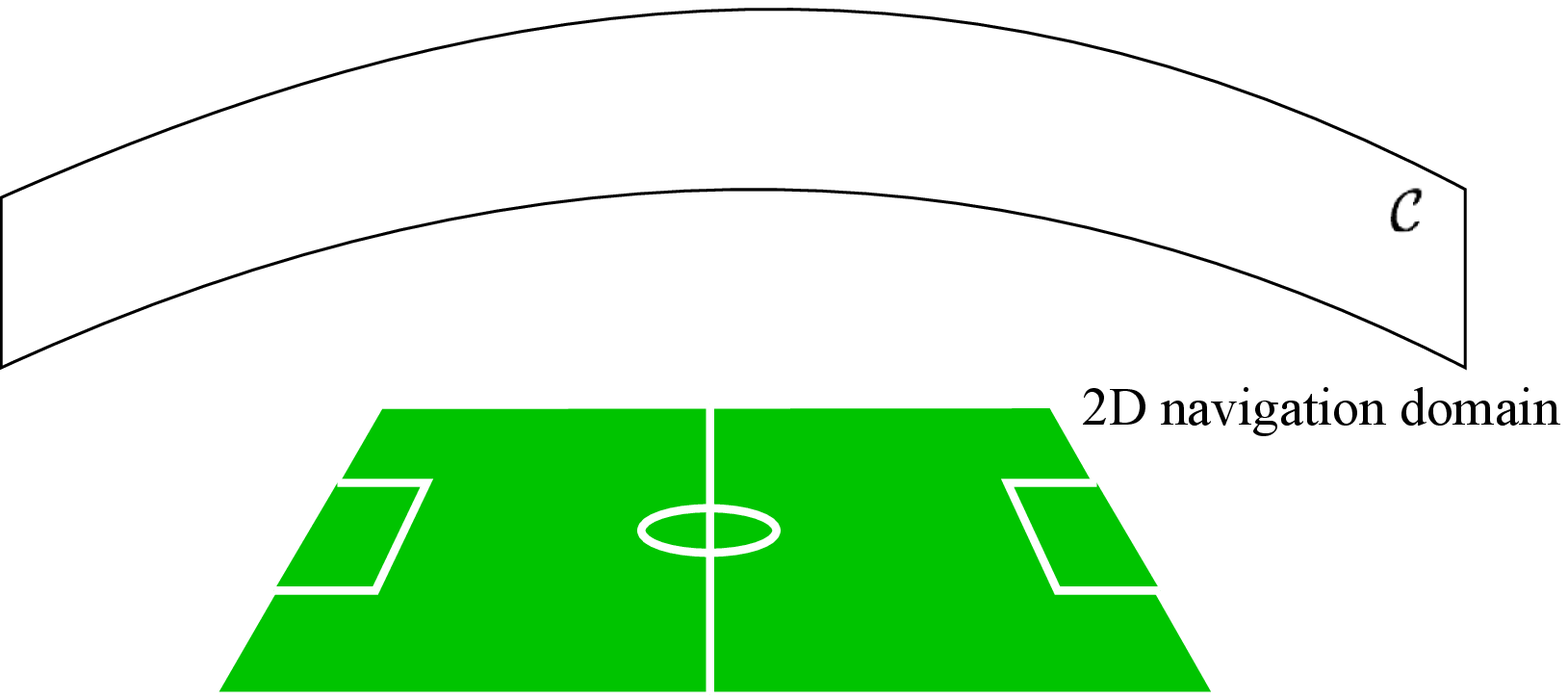,width=6cm}}
\caption{The navigation domain can be 1D or 2D, and is defined by the set of camera parameters $\mathcal{C}$.}
\label{fig:ND}
\end{figure}

We can now formally define the new concept of \emph{navigation domain} as a contiguous region that gathers different viewpoints of the 3D scene $S$, with each of these viewpoints being available for reconstruction as a 2D image by the users (see Fig.~\ref{fig:ND}). This is an alternative to the classical image-based representation used in the literature, where a scene is represented by a set of captured views \cite{Vetro_A_2011_pieee_ove_smvcehms}. In our framework, the concept of captured camera or virtual view does not exist anymore, in the sense that all the images of the navigation domain are equivalent. We denote by $c(X) \in \mathbb{R}^p$ the camera parameter vector associated to the image $X$:
 $$c(X) = c_X =  [\underbrace{ t_x \quad t_y \quad t_z}_{translation} \quad \underbrace{ \theta_x \quad \theta_y \quad \theta_z}_{rotation}]^T.$$ 
 From these parameters, we define the navigation domain as a continuous and bounded domain $\mathcal{C} \in \mathbb{R}^p$. We associate to $\mathcal{C}$ the dual image navigation domain: $\mathcal{X} = \{X | c_X \in \mathcal{C}\}$. In the following, a navigation domain (ND) refers to both the set $\mathcal{C}$ and its dual definition. 

The new concept of navigation domains permits us to have a general formulation of the view switching problem. Naturally it also leads to novel data representation methods. The main idea of our novel approach is first to divide the navigation domain into non-overlapping partitions, $\mathcal{X}_i$, called \emph{navigation segment}. In other words, we have $\mathcal{X} = \bigcup_i \mathcal{X}_i$ with $\mathcal{X}_i \cap \mathcal{X}_j = \emptyset$ for all $i$ and $j$. Then, we represent all the views in one segment with one signal, which is used at decoder for user navigation within the views of the segment.

\begin{figure*}[htb]
  \centering
 \centerline{\epsfig{figure= 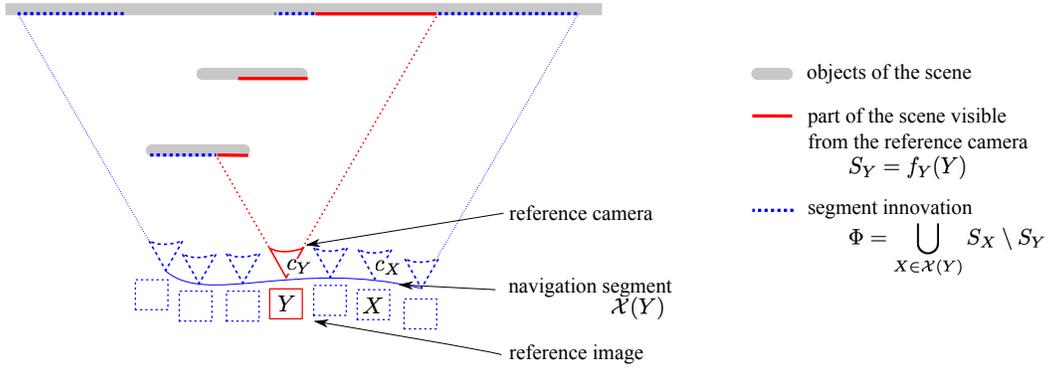,width=14cm}}
\caption{Top-down illustration of the concept of navigation segment for a simple scene with one background (vertical plane) and two foreground objects (vertical rectangles).}
\label{fig:datarepresentation}
\end{figure*}

  Each navigation segment is first described by one \emph{reference} image, called $Y$. This image is important as it is used for the baseline reconstruction of all images in the segment. We thus denote the navigation segment as $\mathcal{X}(Y)$, which represents the set of images that are reconstructed from a reference $Y$ at the decoder. The reference image completely determines the navigation segment under some consideration about the geometry of the scene and the camera positions, as explained later. The part of the scene visible from the reference image $Y$ is called $S_Y = f_Y(Y) $ (it is illustrated in solid lines in Fig.~\ref{fig:datarepresentation}). At the decoder, an image $X$ in the navigation segment is reconstructed using depth-image-based rendering techniques (DIBR \cite{Fehn_C_2004_pspie-sipr_dep_ibrctnatdtv}) that project the frame $Y$  onto $X$,  \emph{i.e.}, the decoder builds $f^{-1}_X(S_Y)$. 
The decoder is however missing the elements of information in $X \setminus f^{-1}_X(S_Y)$ for a complete reconstruction of  each view $X$ of the navigation segment. Since some of these missing elements  in different $X$'s map to the same voxel, we merge the innovation data for different views and define  the global \emph{segment innovation} as
\begin{equation}
\Phi = \bigcup_{X \in \mathcal{X}(Y)} S_X \setminus S_Y.
\end{equation}
It corresponds to a global information that is missing in $Y$ to recover  the whole navigation segment. It is represented in dashed lines in Fig.~\ref{fig:datarepresentation}. The segment innovation $\Phi$ is transmitted to the decoder as auxiliary information that takes the coded form $\varphi = h(\Phi)$.

Equipped with our new data representation method, we can finally describe our communication system in detail. We assume that a server stores the different navigation segments that compose the whole navigation domain. This storage has a general cost $\Gamma$. At the receiver, a user navigates among the views, chooses to build a 2D image $X$ at a viewpoint described by parameters $c_X \in \mathcal{C}$. The only constraint in the navigation is that the user cannot choose randomly his viewpoint, \emph{i.e.,} he has to switch smoothly to the neighboring images. More precisely, the interaction model restricts a user to switch only from view X to a neighboring view Z, where $Z \in neighbor(X)$. The  $neighbor(X)$ corresponds to a set of views that are at a minimum distance of $X$. We thus define a \emph{distance} $\delta$ between two camera parameter vectors $c$ and $c'$ as $\delta:  (c,c')  \rightarrow  \delta(c,c')$. This distance is computed between the camera parameters vectors. We can consider different distances if we want to emphasize rotation or translation in the 3D scene. We note also $\delta:  (X,X')  \rightarrow  \delta(c_X,c_{X'})$ the dual distance between two images $X$ and $X'$.  Since $\mathcal{C}$ is a continuous set, we define $\Delta$ as the \emph{navigation step}, which corresponds to the distance between two different images chosen at consecutive instant.
We assume that the user can send its position in the navigation domain every $N_T$ frame\footnote{If $f$ is the frame rate, $N_T$ can be expressed in seconds by dividing the value expressed in number of frames by $f$}. Once the user sends its position, the server transmits all the navigation segments that the user might need in the next $N_T$ instants. We define the \emph{navigation ball} as the set of achievable viewpoints in the next $N_T$ instants from the viewpoint $X$ as:
\begin{equation}
B(X,N_T\Delta) = \{X'\in\mathcal{X} | \delta(X,X')<N_T\Delta \}.
\label{eq:navBall}
\end{equation}
In other words, the server sends all navigation segments $\mathcal{X}(Y_i)$ such that $\mathcal{X}(Y_i) \cap B(X,N_T\Delta) \neq \emptyset$.
Finally, the user navigation depends on the \emph{a priori} view popularity distribution, $p(X)$ (with $X\in \mathcal{X}$), which corresponds to a dense probability distribution over the views. It describes the relative popularity of the viewpoints, with $\int_{X\in \mathcal{X}} p(X) = 1$, and captures the fact that  all viewpoints do not have the same probability to be reconstructed at decoder in practice.

\subsection{Data coding}\label{sec:implementation}
\begin{figure}[htb]
  \centering
 \centerline{\epsfig{figure= 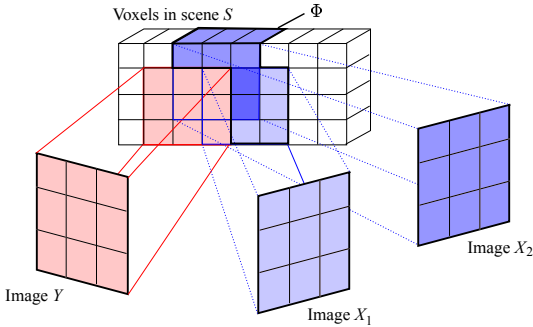,width=7cm}}
\caption{Example of $\Phi$ construction, when images $Y$, $X_1$ and $X_2$ are projected to the 3D scene. In that example, $\Phi$ is made of $9$ voxels, thus $|\Phi| = 9$.}
\label{fig:phi}
\end{figure}

After we have derived a representation of the static 3D scene, we finally describe how the chosen representation can be efficiently encoded using existing coding tools.  Recall that each navigation segment is composed of one reference image $Y$ and some auxiliary information $\varphi = h(\Phi)$, where $\Phi$ is the segment innovation. First, the images $Y$ (color and depth data) are coded and stored using classical intra frame codecs such as H.264/AVC Intra \cite{Wiegand_T_2003_tcsvt_ove_hvcs}. We use such reference images to generate all the other views of the navigation segment $\mathcal{X}(Y)$ \emph{via} view synthesis. As explained before, the set of frames $X\in\mathcal{X}(Y) \setminus Y$ contains a certain innovation $\Phi$ that represents the global novelty of the views in the navigation segment with respect to $Y$. In practice, we estimate this set as follows (see in Fig.~\ref{fig:phi}). We first project the image $Y$ in the 3D scene using depth information (in other words, we compute $S_Y$). Then, we project every frame $X$ from the segment $\mathcal{X}(Y)$ into the 3D scene using depth information (\emph{i.e.}, we compute $S_X$). In our representation, each pixel is associated with a voxel in the 3D space and  voxels can be shared by two images. In practice, $\Phi$ 
is the union of voxels visible in views in $\mathcal{X}(Y)$ but not visible in $Y$. In order to avoid redundancies, the  voxels shared by different views in $\mathcal{X}(Y)$ are only represented once in $\Phi$. In the following, we will use the concept of size of $\Phi$, which simply corresponds to the number of voxels in the set $\Phi$, denoted as $|\Phi|$. We will see that this size has a strong impact on the rate of the coded auxiliary information $\varphi = h(\Phi)$, denoted by $|\varphi|$. 

We still have to encode the auxiliary information to reduce its size. 
One way of doing it is to first project the innovation set onto a well chosen viewpoint, \emph{i.e.,} a viewpoint where the voxels do not overlap and which can gather the whole segment innovation $\Phi$. We remark that this is generally possible when the cameras are aligned. 
The encoding function (\emph{i.e.}, the function $\varphi = h(\Phi)$) then consists in building a quantized version of DCT blocks from this projected innovation image. The innovation segment image is thus divided into small pixel blocks that are DCT transformed and quantized\footnote{The quantization step applied here are, for the moment, chosen empirically so that we reach a similar quality in the occlusion than in the rest of the image.}. The bitstream is then encoded with a classical arithmetic coder. This method is not fully optimized in terms of compression and is certainly neither exclusive nor unique; it however nicely fits the design choices described above. If the navigation domain is more complex, our approach can be extended to the layered depth image (LDI \cite{Shade_J_1998_psiggraph_lay_di}) format, to deal with voxels overlapping. In that case, auxiliary information in  each layer can be DCT transformed and quantized. We outline here that the design of the auxiliary information coding technique does not depend on the decoder.

At the decoder we exploit the auxiliary information in a reconstruction strategy that is based on the Criminisi's inpainting algorithm \cite{Criminisi_A_2004_tip_reg_forebii}. The first step of the inpainting algorithm chooses the missing image patch that has the highest priority based on image gradient considerations. A second step then fills in the missing information by using a similar patch from the reconstructed parts of the image. We modify the original  Criminisi's inpainting algorithm by introducing  in this second step a distance estimation between the candidate patch and the auxiliary information in the navigation segment. The hole-filling technique thus chooses a patch that corresponds to the auxiliary information $h(\Phi)$, more exactly to its projected version onto the current viewpoint. Finally, it  is important to note that the reconstruction technique is independent of the type of hash information that is transmitted.

The impact of the reference data compression is twofold. First, it induces some error propagation in the texture of the synthesized frames, due to the fact that part of the synthesized image takes its information from the reference view. We leave this issue for future work, but it certainly deserves careful attention in the design of more evolved coding strategies.  Second, the compression of the reference image influences the representation (reference + auxiliary information) itself. This is only linked with the compression of the depth image associated to the reference. Changes in the depth map lead to different innovation and thus different auxiliary information. In our tests, we have seen that such a phenomenon could be very important but only when depth images are coarsely compressed. This is why we mostly consider high quality depth maps in our framework.

\section{Optimal partitioning of the navigation domain}\label{sec:partitioning}

\subsection{Constrained partitioning}
The new data representation proposed above raises an important question, namely the effective design of the navigation segments. We show here how the partitioning of the navigation domain into navigation segments can be optimized under rate and storage constraints.
We can describe the navigation domain as the union of $N_V$ navigation segments by $ \mathcal{X} = \bigcup_{i=1}^{N_V} \mathcal{X}(Y_i)$, where $\mathcal{X}(Y_i)$ is the set of images reconstructed from the reference image $Y_i$ and the associated auxiliary information in a navigation segment.

\begin{figure}[htb]
  \centering
\begin{minipage}{0.48\linewidth}
 \centerline{\epsfig{figure= 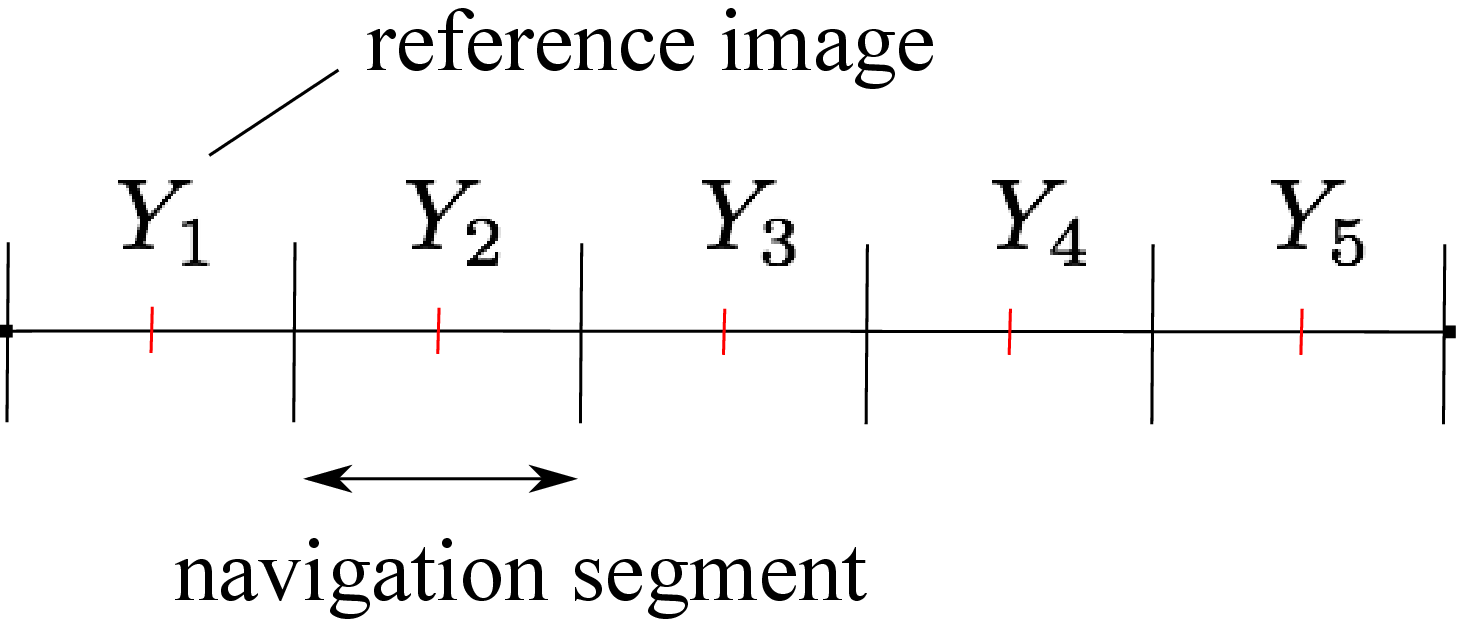,width=4cm}}
\centerline{(a) distance-based}
\end{minipage}
\begin{minipage}{0.48\linewidth}
 \centerline{\epsfig{figure= 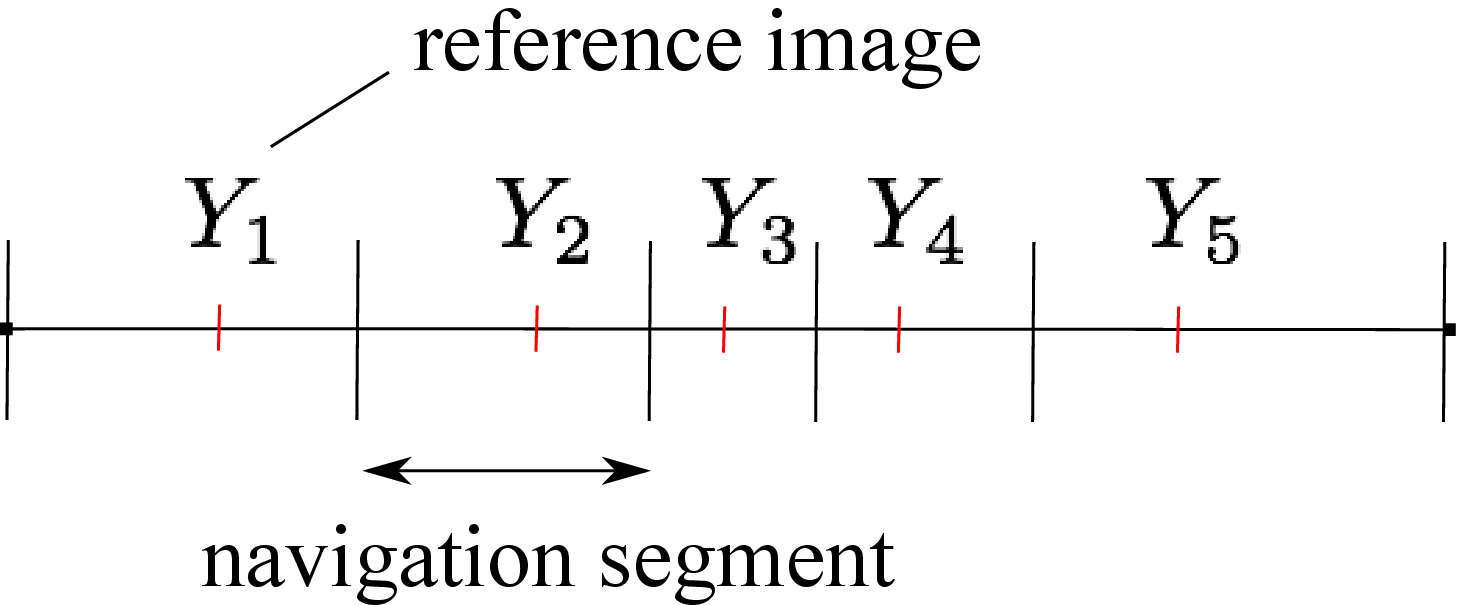,width=4cm}}
\centerline{(b) similarity-based}
\end{minipage}
\caption{Illustration of the difference between the  distance-based and similarity-based partitioning for 1D navigation domain. }
\label{fig:frameRepEx}
\end{figure}

\begin{figure}[htb]
  \centering
\begin{minipage}{0.50\linewidth}
 \centerline{\epsfig{figure= 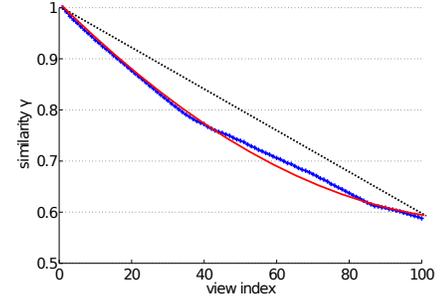,width=6cm}}
\centerline{(a) similarity with respect to view 1}
\end{minipage}
\begin{minipage}{0.50\linewidth}
 \centerline{\epsfig{figure= 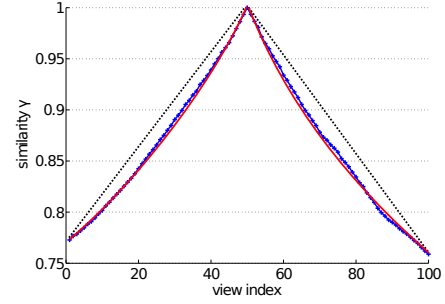,width=6cm}}
\centerline{(b) similarity with respect to view 50}
\end{minipage}
\caption{Similarity evolution (blue crosses) in function of the view index of a navigation domain, in which the images are equidistant. Black dashed line corresponds to linear interpolation between the extreme values while red plain curve is a non-linear interpolation of the curve which obviously fits better the similarity evolution.}
\label{fig:similarityVSdistance}
\end{figure}

Let us first study a simple scenario, which permits to define the new concept of similarity between two frames. We assume that  $N_V$ is given and that the reference images $Y_i$'s are already fixed. A natural way of defining a navigation segment $\mathcal{X}(Y)$ consists in decomposing the ND based on the distance between cameras:
\begin{eqnarray}
&&\forall i\in [1,N_V],\\
&& \quad \mathcal{X}(Y_i) = \{ Y_i\} \cup \{ X \in \mathcal{X}  |  \forall j \neq i,  \delta(X,Y_i) \leq \delta(X,Y_j) \}. \nonumber 
\end{eqnarray}
This definition leads to equidistant reference image distribution over the navigation domain as shown in Fig.~\ref{fig:frameRepEx}(a). However, this definition takes into account neither the scene characteristics nor the innovation between the images. In order to take into account the scene information in the partitioning process, we  define \emph{geometrical similarity} $\gamma$ between two images as:
\begin{equation}
\gamma: \quad (X,X') \quad \rightarrow \quad \gamma(X,X') = \left| S_X \cap S_{X'} \right|.
\label{eq:simi}
\end{equation}
For the sake of conciseness, the term geometrical similarity will be replaced by similarity in the rest of the paper. This similarity definition lays the foundation of a new kind of correlation between images that share a set of identical pixels but also contain sets of independent pixels. In other words, instead of considering a model where the correlation between two images is an error all over the pixels, as it is classically adopted in image coding, we use here a model where two pixels in different images correspond or not to the same voxel in the 3D scene (\emph{i.e.}, they are either equal or totally independent in the projected images). This new kind of correlation between images is measured by the similarity function of Eq.~(\ref{eq:simi}). This leads to a novel partitioning strategy defined as:
\begin{eqnarray}
&&\forall i\in [1,N_V], \\
&&\quad \mathcal{X}(Y_i) = \{ Y_i\} \cup \{ X \in \mathcal{X}  |  \forall j \neq i,  \gamma(X,Y_i) \geq \gamma(X,Y_j) \}. \nonumber
\label{eq:neighborhood}
\end{eqnarray}
Interestingly, this solution depends on the quantity of innovation between two images and leads to non-equidistant partitioning. Typically, the navigation segments are smaller if the similarity varies quickly with the distance between cameras (Fig.~\ref{fig:frameRepEx}(b)).
To illustrate the fact that similarity is not linearly dependent on the distance between cameras, we present a simple experiment in Fig.~\ref{fig:similarityVSdistance}. For the \emph{Ballet} sequence \cite{web_microsoft_ballet_break}, we build a navigation domain made of $100$ equidistant viewpoints. For two reference images (index 1 in Fig.~\ref{fig:similarityVSdistance}(a) and 50 in Fig.~\ref{fig:similarityVSdistance}(b)) we calculate their similarity with all the other frames of the navigation domain. The similarity is expressed here between 0 and 1 and corresponds to a percentage of common pixels\footnote{The  similarity is normally defined as a number of voxels in the 3D space, however, for this test, we have chosen to divide it by the size of the image in order to obtain a value between 0 and 1, which makes the interpretation easier.}, \emph{i.e.,} the number of pixels that are associated to the same voxels in the 3D scene. We can actually see that the evolution of the similarity function is not linear with the view index nor with the distance; the non-linear (plain  lines) interpolation function fits better  the similarity function than the linear one (dashed  line).

Equipped with this new fundamental notion of similarity, let us develop further our framework towards optimal partitioning of the navigation domain. This optimal partitioning is obtained by fixing the right number of reference views $N_V^*$ and choosing the proper reference images $Y_i^*$. Partitioning is optimized with respect to a storage size $\Gamma$, which corresponds to the total cost of storing all the navigation segments, and with respect to a rate $R$, which corresponds to an average transmission cost. We assume that the navigation step $\Delta$ that correspond to the distance between two consecutive images in the navigation is fixed. We further assume that if a user starts its navigation on a reference frame, the navigation segment is sufficiently big to enable independent navigation during $N_T$ time instant without transmission of another navigation segment. The optimal partitioning problem consists in defining the set of partitions (\emph{i.e.}, the number of navigation segments $N_V$ and the reference images $Y_i$'s) that minimize the streaming rate $R(N_V, \{Y_i\})$, while the total storage $\Gamma(N_V, \{Y_i\})$ is smaller than a storage capacity $\Gamma_{max}$. Formally, it can be posed as:
\begin{eqnarray} 
 &(N_V^*, \{ Y_i^*\})  = \argmin_{(N_V, \{ Y_i\})} R(N_V, \{ Y_i\}) ) \\
 &\mbox{under the constraint that} \  \Gamma(N_V, \{ Y_i\}) \leq \Gamma_{\rm max}. \nonumber \label{eq:minimization} 
  \end{eqnarray}
  We  rewrite the above as an  unconstrained problem with help of a Lagrangian multiplier $\lambda$ as
  \begin{eqnarray}
 (N_V^*, \{ Y_i^*\})  = \argmin_{(N_V, \{ Y_i\})} R(N_V, \{ Y_i\}) ) + \lambda  \Gamma(N_V, \{ Y_i\}). \label{eq:minimization_Lag}
 \end{eqnarray}
The storage $\Gamma(N_V, \{Y_i\})$ depends on both the size of the reference frame $|Y_i|$ and  the auxiliary information $|\varphi_i|$, with $\varphi = h(\Phi)$ being the coding function for each navigation segment. It can be formulated as $\Gamma(N_V, \{Y_i\}) = \sum_{i=1}^{N_V} S(\mathcal{X}(Y_i))$, where $S (\mathcal{X}(Y_i)) = \left(|Y_i| + |\varphi_i| \right)$ is the size of a navigation segment. The transmission rate $R$ corresponds to the expected size of the information to be sent after each request and is driven by the  size of the navigation segments. Note that, formally, it differs from the classical definition of transmission rate expressed in bit per second.  It depends on navigation models or view popularity and is written as:
\begin{eqnarray}
 R(N_V, \{Y_i\}) &=& \sum_{i=1}^{N_V} P(\mathcal{X}(Y_i))  S(\mathcal{X}(Y_i)) \nonumber \\
 &=& \sum_{i=1}^{N_V} P(\mathcal{X}(Y_i)) (|Y_i| + |\varphi_i|)\label{eq:rate}
 \end{eqnarray}
where  $P(\mathcal{X}(Y_i)) = \int_{X\in\mathcal{X}(Y)} p(X) $ corresponds to the probability that the user navigates in  segment $\mathcal{X}(Y)$, as proposed in Sec.~\ref{sec:framework}. We propose below a method to solve the optimal partitioning problem of Eq.~(\ref{eq:minimization_Lag}).

\subsection{Optimization method}\label{sec:partOpt}

We first assume that the number of segments $N_V^*$ is given. In this case, we notice that the optimization problem of Eq.~(\ref{eq:minimization_Lag}) is similar to a problem of vector quantization \cite{Gersho_A_1992_vec_qsc}. The vector quantization problem consists in dividing the vector space in different partitions, represented by codewords that are chosen to minimize the reconstruction distortion under a rate constraint. Here we want to find a partitioning of the navigation domain that minimizes the rate under a storage constraint, while the quality of the reconstruction is not affected. In Lloyd algorithm \cite{Gersho_A_1992_vec_qsc} for vector quantization, the positions of the codeword determine the quantization cells; similarly, the position of $Y_i$ determines the navigation segment $\mathcal{X}(Y_i)$ in our partitioning problem. More precisely, in an ideal case, the definition of the navigation segments becomes 
$\mathcal{X}(Y_i) =  \argmin_{\{X\}} | \varphi_i |$ when the $Y_i$ are fixed and when $\bigcup_i \mathcal{X}(Y_i)$ covers the whole navigation domain $\mathcal{X}$.
We consider that it can be achieved from Eq.~(\ref{eq:neighborhood}), which builds the segment with the elements that have a higher similarity with the reference frame than with  the reference frames of the other navigation segments. The problem now consists in selecting the reference frames $Y_i$'s. We consider a simple iterative algorithm that performs three steps:
\begin{itemize}
\item step 1: initialize the reference frames $Y_i$'s at equidistant positions in order to avoid local minima.
\item step 2: derive the optimal navigation segments given the reference frames $Y_i$'s, based on frame similarity criteria in Eq.~(\ref{eq:neighborhood}).
\item step 3: refine the reference frame in each navigation segment in order to minimize storage and rate costs in Eq.~(\ref{eq:minimization_Lag}).
\end{itemize}
The algorithm then proceeds iteratively and alternates between steps 2 and 3. It terminates when the refinement in step 3 does not provide a significant storage and rate gain. While global optimality cannot be guaranteed in this family of alternating algorithms, the convergence is guaranteed, because the same objective function $R + \lambda \Gamma$ is minimized in both steps 2 and 3, and the objective function is bounded from below.

It remains now to define the optimal number of segments, \emph{i.e.}, the value $N_V^*$. For that purpose, we need to define a maximum number of navigation segments $M$. It corresponds to the case where all the segments have the minimum acceptable area, \emph{i.e.}, the area of the navigation ball $B(X,N_T\Delta)$ defined in Eq~(\ref{eq:navBall}). We write $M$ as follows :
$$ M = \frac{area(ND)}{area(B(X,N_T\Delta))} .$$
The $area$ of $C \subset \mathcal{C}$ is defined as $\int_{x\in \mathcal{C}}\mathds{1}(x)dx$ (where $\mathds{1}$ is the classical indicator function). It results that $N_V^*$ lies between 1 and $M$. Ideally, we may determine $N_V^*$ with a similar formulation than before, as

\begin{equation}
 N_V^* = \argmin_{1 \leq N_V \leq M} \Gamma (N_V, \{ Y_i\})  + \mu R_{\rm max}(N_V, \{ Y_i\}) \label{eq:numberOfPartitions}
\end{equation}
where $R_{\rm max}$ is the maximum navigation segment size  that the user receives per request during navigation. The parameter $\mu$ regulates the relative importance of the rate with respect to the storage cost. In practice, to solve Eq.~(\ref{eq:numberOfPartitions}), we neglect the influence of $\{ Y_i\}$ and we estimate the storage $\bar{\Gamma}$ and rates $\bar{R}_{\rm max}$  values at a high level:
\begin{itemize}
\item $\bar{\Gamma} = N_V \bar{|Y|} + N_V \bar{|\varphi|} $, where $\bar{|Y|}$ and $ \bar{|\varphi|}$ are estimations of the average reference frame rate and reference auxiliary information rate. They are deduced from the coding strategy adopted for $\varphi = h(\Phi)$.
\item $\bar{R}_{\rm max} = \bar{|Y|} +  \bar{|\varphi|}.$
\end{itemize}

Finally we have the optimal value of the number of navigation segments by exhaustive search of $N_V \in [1,M]$, as
\begin{equation}
 N_V^* = \argmin_{1 \leq N_V \leq M} (N_V + \mu) \bar{|Y|} + (N_V + \mu) \bar{|\varphi|}. 
 \label{eq:NV}
 \end{equation}

\section{Experiments}\label{sec:exp}

\subsection{Setup}\label{sec:setup}
\begin{figure}[htb]
  \centering
\begin{minipage}{0.32\linewidth}
 \centerline{\epsfig{figure= 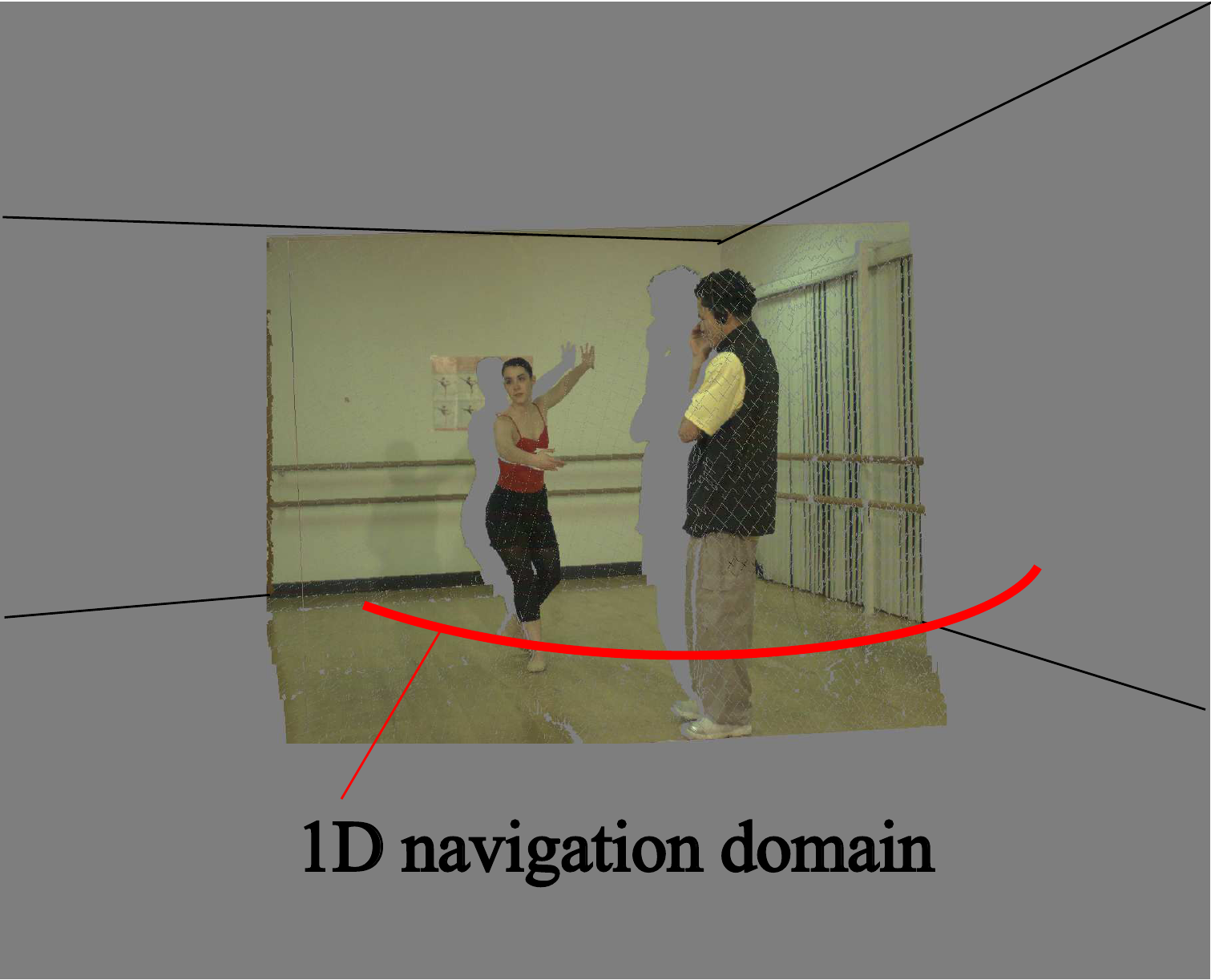,width=2.8cm}}
\end{minipage}
\begin{minipage}{0.32\linewidth}
 \centerline{\epsfig{figure= 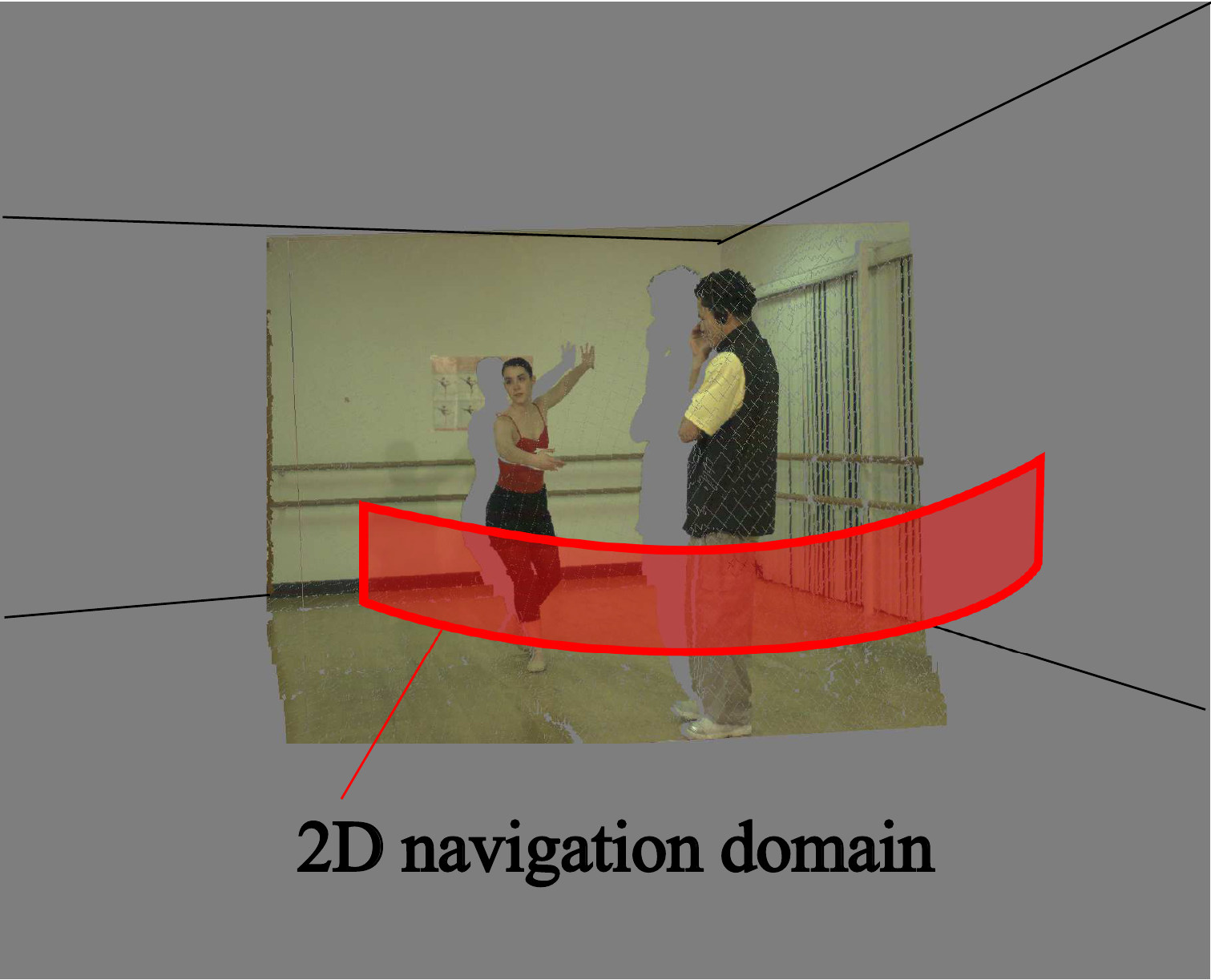,width=2.8cm}}
\end{minipage}
\begin{minipage}{0.32\linewidth}
 \centerline{\epsfig{figure= 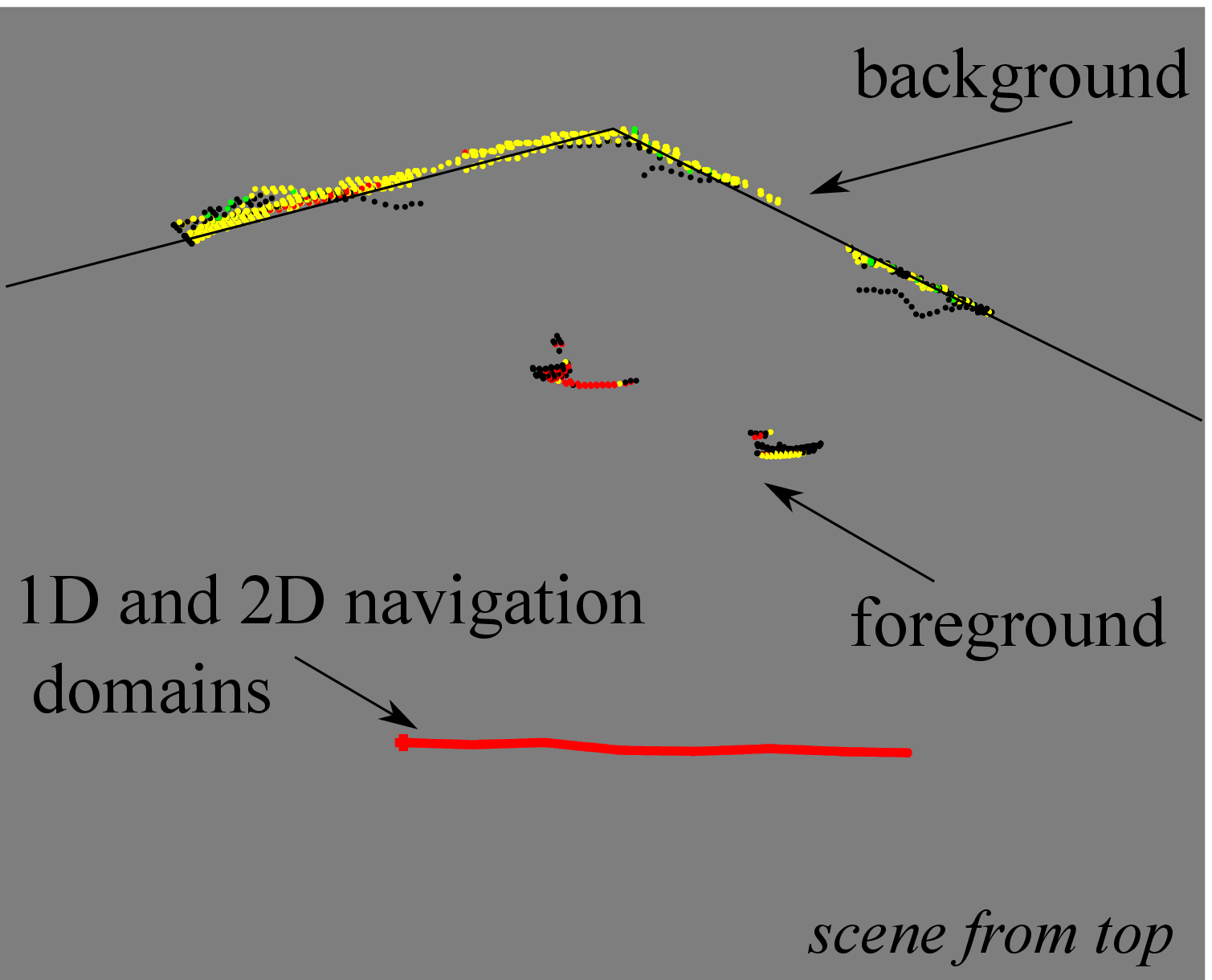,width=2.8cm}}
\end{minipage}
\caption{Illustration of 1D and 2D navigation domains used in the experiments for \emph{Ballet} sequence, and of a top view of the 3D scene.}
\label{fig:ballet_break_ND}
\end{figure}

Our novel interactive system is tested on two well-known multiview sequences provided by Microsoft research \cite{web_microsoft_ballet_break}\footnote{Since we are considering the navigation in a static scene, we only consider frames captured at time $1$.}, namely \emph{Ballet} and \emph{Breakdancer}.  Each of these sequences is composed of eight texture and depth videos and their associated intrinsic and extrinsic parameters. From these multiview images, we build a navigation domain that is composed of $120$ viewpoints (texture and depth), as illustrated in Fig.~\ref{fig:ballet_break_ND} for \emph{Ballet} sequence. We also build a 2D navigation domain that consists of $5$ distinctive rows of $120$ horizontal aligned viewpoints. In order to create the viewpoints that are not present in the original sequences, we use view synthesis techniques \cite{Muller_K_2011_pieee_tdv_rudm}. All of the images (camera images and synthetic images) form our input dataset; they are considered as original images, and  can be chosen as reference frames by the partitioning algorithm. Finally, we index images in this set of equidistant viewpoints from $1$ to $120$ for the 1D navigation domain, and from $(1,1)$ to $(5,120)$ for the 2D one.

\subsection{Disocclusion filling based on coded auxiliary information}\label{sec:disfill}
\label{sec:exp_inp}

\begin{figure*}[htb]
  \centering
\begin{minipage}{0.3\linewidth}
 \centerline{\epsfig{figure= 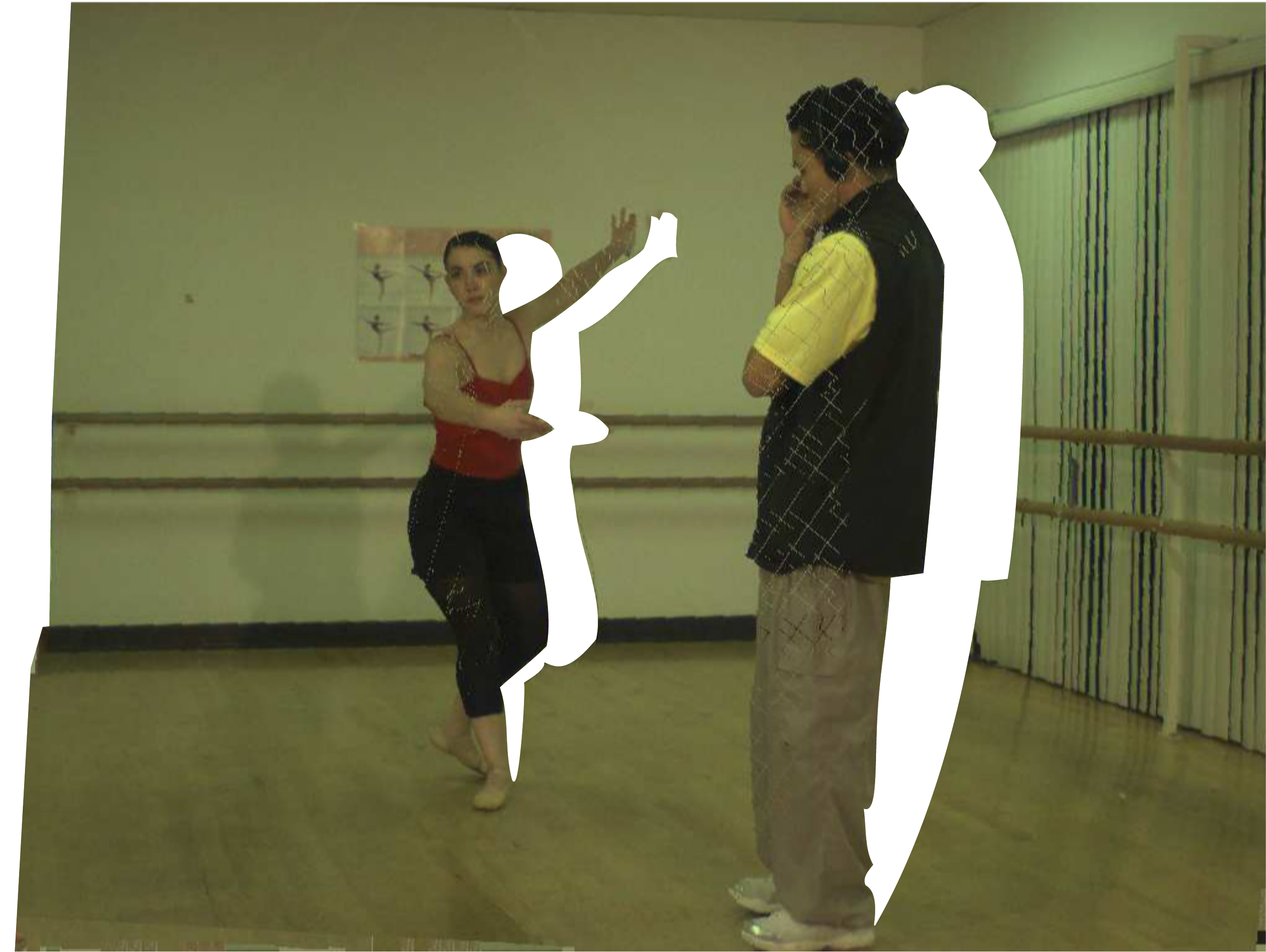,width=4cm}}
\centerline{\begin{minipage}{0.8\linewidth}
\small (a) reference image projection on X ($f^{-1}_X(S_Y)$)\\
\end{minipage}}
\end{minipage}
\begin{minipage}{0.3\linewidth}
 \centerline{\epsfig{figure= 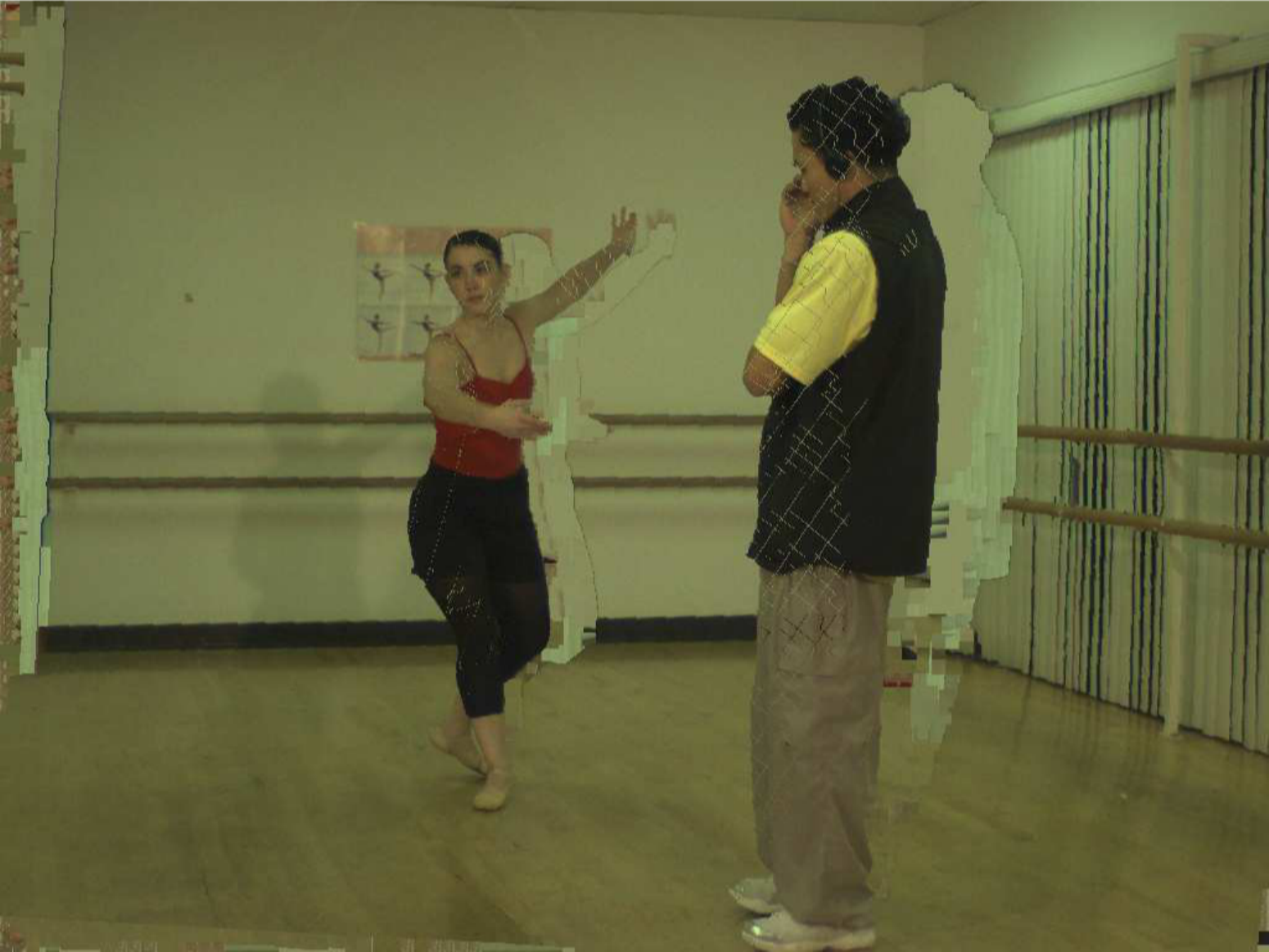,width=4cm}}
\centerline{\begin{minipage}{0.8\linewidth}
\small (b) $X \setminus f^{-1}_X(S_Y)$ recovery without auxiliary information $\varphi$
\end{minipage}}
\end{minipage}
\begin{minipage}{0.3\linewidth}
 \centerline{\epsfig{figure= 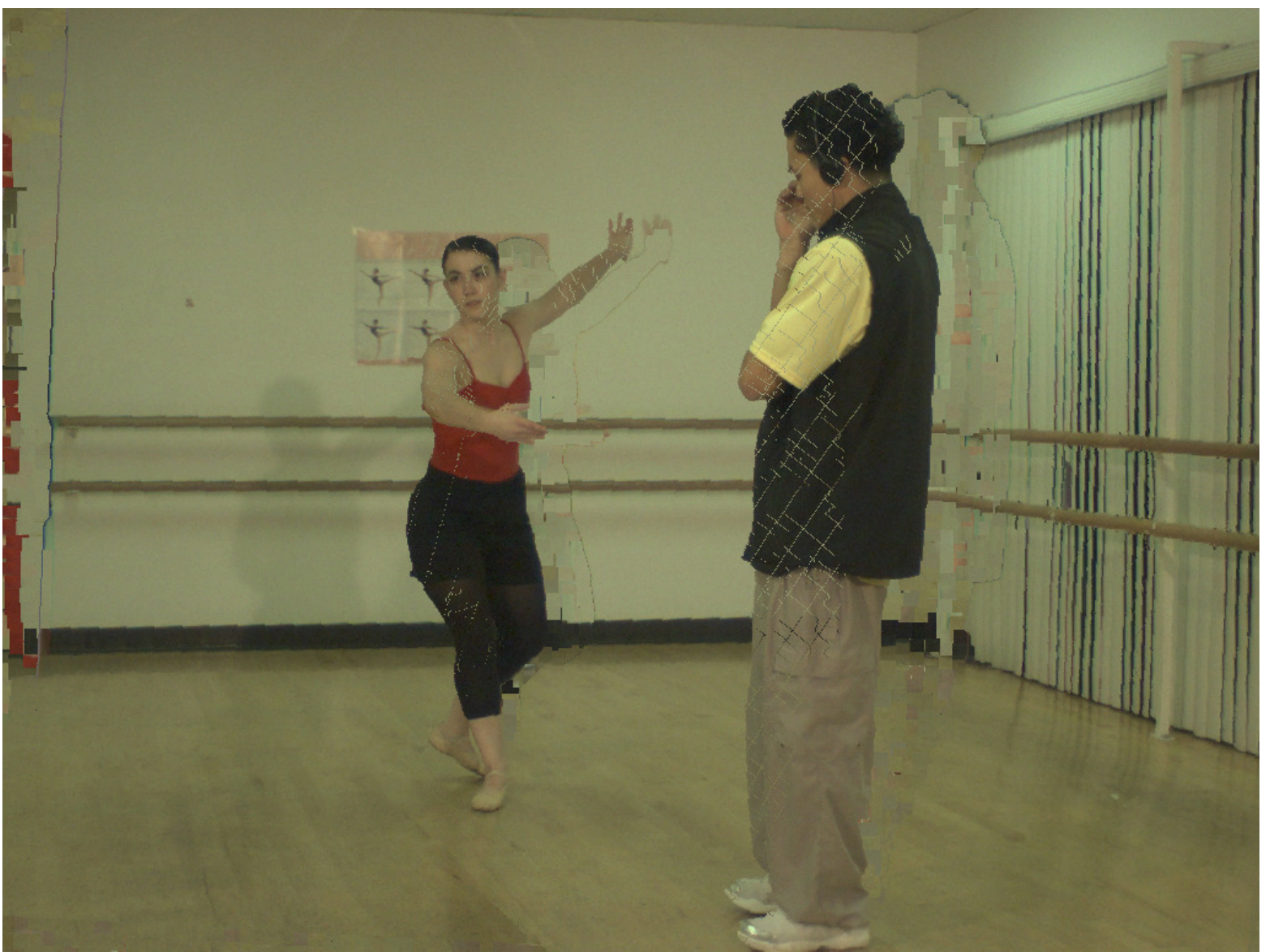,width=4cm}}
\centerline{\begin{minipage}{0.8\linewidth}
\small (c) $X \setminus f^{-1}_X(S_Y)$ recovery with DCT coded auxiliary information $\varphi$
\end{minipage}}
\end{minipage}
\caption{Visual results for  reconstruction of view 2 in \emph{Ballet} using view $1$ as reference image. We compare classical inpainting method (b) and proposed guided inpainting method (c).}
\label{fig:inpVis}
\end{figure*}

\begin{figure}[htb]
\centerline{\epsfig{figure= 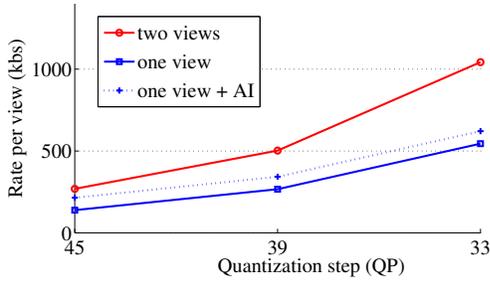,width=7cm}}
\caption{Rate comparison between different representations of the navigation segment: single view, two views, one reference image + auxiliary information. The rendering quality at each quantization step is similar in the three representations.}
\label{fig:inpVis_rate}
\end{figure}

We first study the performance of disocclusion filling algorithm based on auxiliary information. This permits to validate the reconstruction strategy that is at the core of our new data representation method.
An example of reconstruction with the proposed inpainting technique is shown in Fig.~\ref{fig:inpVis}(a). It is obtained by first projecting a reference view $Y$ on a virtual view $X$ (see Fig.~\ref{fig:inpVis}(a), the disocclusions are in white). Second, the disoccluded regions are reconstructed using the classical Criminisi's algorithm (Fig.~\ref{fig:inpVis}(b)) and our guided inpainting method (Fig.~\ref{fig:inpVis}(c)). We can see that the reconstructed quality obtained with our method is very satisfying. Moreover, the side information used for this illustrative example is not heavy in terms of bitrate, as shown in Fig.~\ref{fig:inpVis_rate}. In these experiments, we measure the rate (at different quantization steps for both the reference and the auxiliary information coding) of the following schemes: a single view transmission, two views coded jointly, and our proposed representation (one view and the auxiliary information $\varphi$). We observe that the rate of our representation method is much smaller than the rate needed for sending two reference views for synthesis.

\begin{figure}[htb]
\centerline{\epsfig{figure= 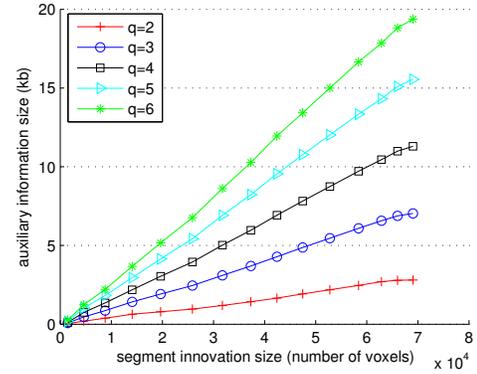,width=7cm}}
\caption{Illustration of the evolution of the size of auxiliary information  $|\varphi|$  as a function of the number of voxels in the segment innovation $|\Phi|$; the auxiliary information is coded with a DCT-based scheme with uniform quantization of the coefficients, where $q$ corresponds to the number of bits used to describe each DCT coefficient.}
\label{fig:icipPhiandVarPhi}
\end{figure}

Recall now that the rate and storage costs depend on $|\varphi|$, which is the size (in kbits) of the auxiliary information (see Eq. (\ref{eq:minimization_Lag})). This auxiliary information is a compressed version of the segment innovation $\Phi$. The quantity of information in $\varphi$ is increasing when the number of elements in the segment innovation $\Phi$ is increasing. We can also observe that the increase is almost linear with the auxiliary information design presented above (see Fig.~\ref{fig:icipPhiandVarPhi}). 
Hence, we have chosen to present the next performance results in terms of number of voxels in the segment innovation $|\Phi_i|$ instead of rate and the storage costs. This advantageously leads to  presenting general results that can be adapted to any kind of encoding function $h$. We will however show  in Sec.~\ref{sec:rdeval} some rate and storage results obtained with a practical implementation of the system (based on a auxiliary information constructed using DCT coefficients as introduced in Sec.~\ref{sec:exp_inp}).

\begin{figure}[htb]
  \centering
\begin{minipage}{0.6\linewidth}
 \centerline{\epsfig{figure= 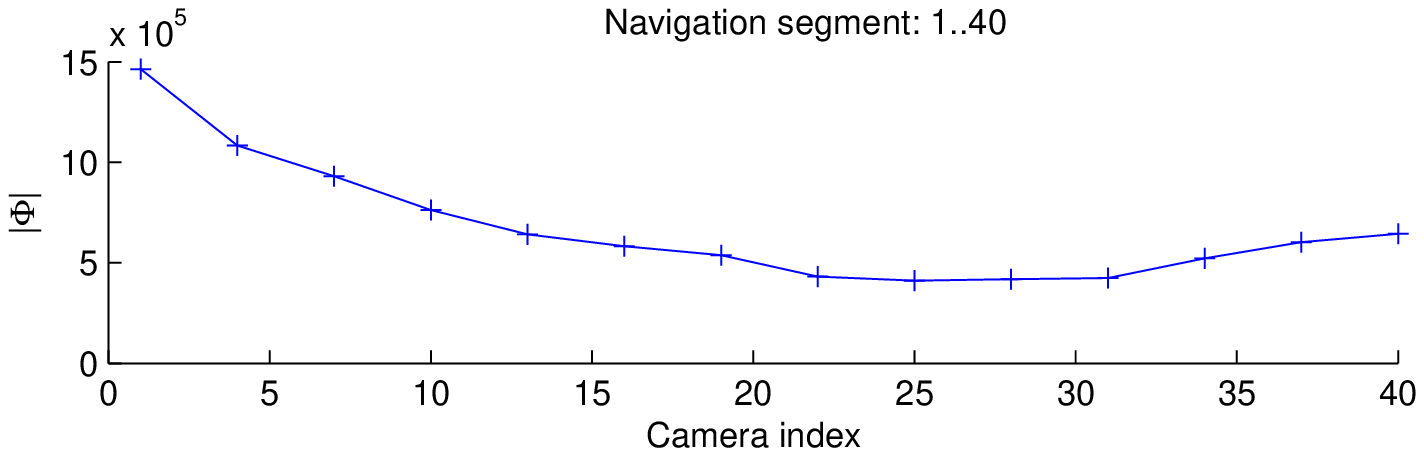,width=8cm}}
\vspace{0.1cm}
\end{minipage}
\begin{minipage}{0.6\linewidth}
 \centerline{\epsfig{figure= 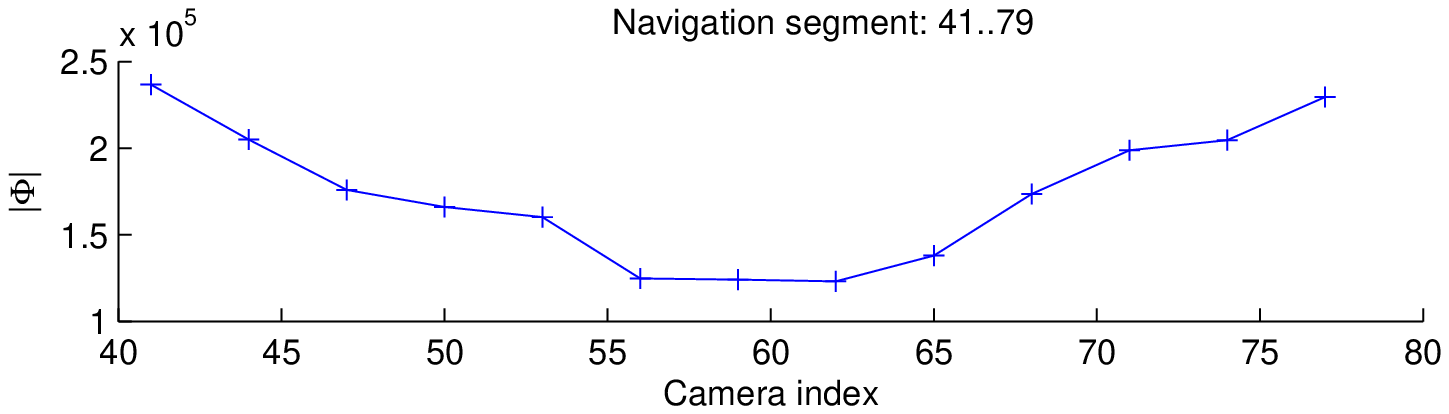,width=8cm}}
\vspace{0.1cm}
\end{minipage}
\begin{minipage}{0.6\linewidth}
 \centerline{\epsfig{figure= 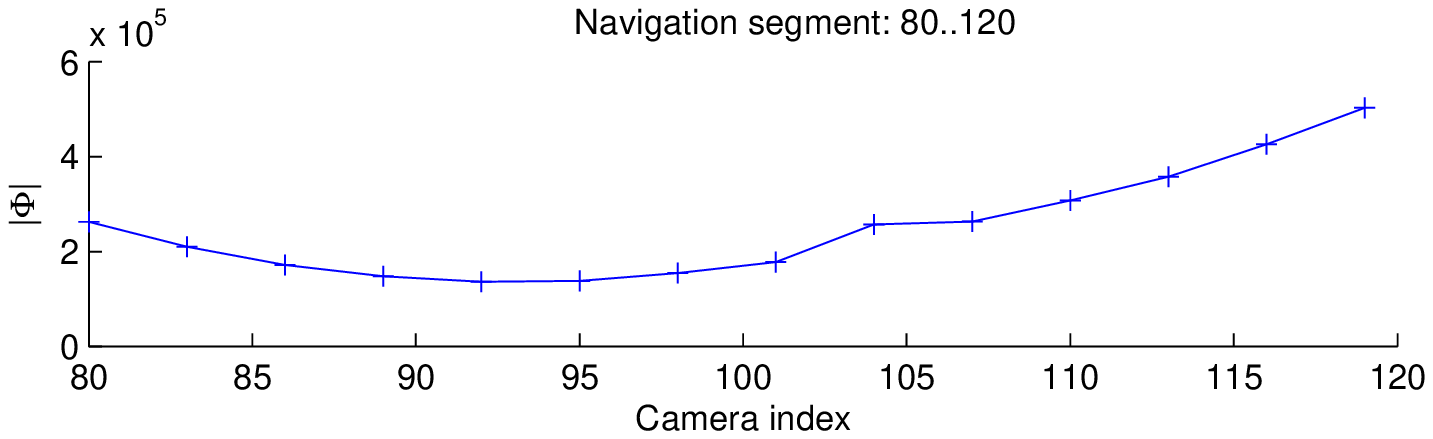,width=8cm}}
\end{minipage}
\caption{Size of the segment innovation $\Phi$ (measured in number of voxels) for \emph{Ballet} sequence, as a function of the reference frame position (expressed in terms of camera index within the general navigation domain) for a 1D navigation domain and fixed navigation segments.}
\label{fig:sizePhi}
\end{figure}

\begin{figure}[htb]
  \centering
\begin{minipage}{0.6\linewidth}
 \centerline{\epsfig{figure= 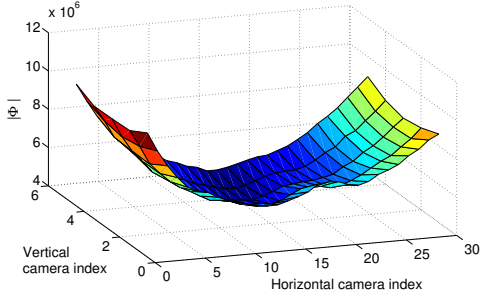,width=7cm}}
\end{minipage}
\caption{Size of the segment innovation $\Phi$ (measured in number of voxels) for \emph{Ballet} sequence, as a function of the reference frame position (expressed in terms of camera index within the general navigation domain) for a 2D navigation domain and fixed navigation segments.}
\label{fig:sizePhi2D}
\end{figure}

\subsection{Influence of reference view}\label{sec:refview}
We now study the influence of the position of a reference view within a navigation segment. One of the strengths of the proposed representation is to avoid the differentiation between captured and synthesized views. Every frame is considered with the same importance, which gives a new degree of freedom  in navigation performance optimization via proper selection of the reference view $Y_i$.
We evaluate the impact of the position of $Y_i$ on the size of the segment innovation $\Phi$. We illustrate in Fig.~\ref{fig:sizePhi} and \ref{fig:sizePhi2D} the typical evolution of $\Phi$ as a function of $Y_i$, in 1D and 2D navigation domains. More precisely, we fix the navigation segments and vary the position of the reference frames $\{Y_i\}$. For each position, we calculate $|\Phi|$ as explained in Sec.~\ref{sec:implementation}. We see that the evolution of the segment innovation size is approximately convex, but non regular and non symmetric. The size of the auxiliary information clearly depends on the scene content. We see that the position of $Y_i$ has a strong impact on the size of the segment innovation, and therefore on the rate of the encoded auxiliary information. We see that the size $|\Phi|$ can even vary in a ratio of 1:2, depending on the position of the reference view.

\subsection{Optimal partitioning}\label{sec:optpart}

\begin{figure}[htb]
  \centering
\begin{minipage}{0.48\linewidth}
 \centerline{\epsfig{figure= 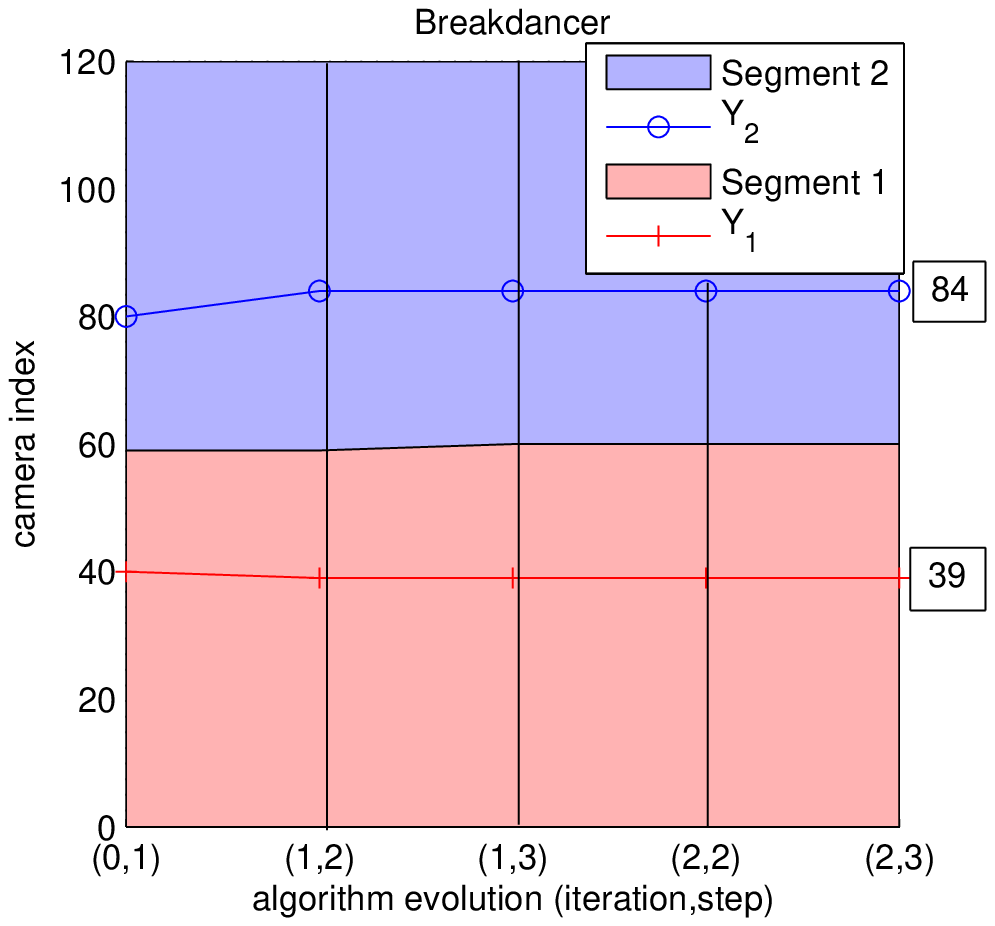,width=4cm}}
\end{minipage}
\begin{minipage}{0.48\linewidth}
 \centerline{\epsfig{figure= 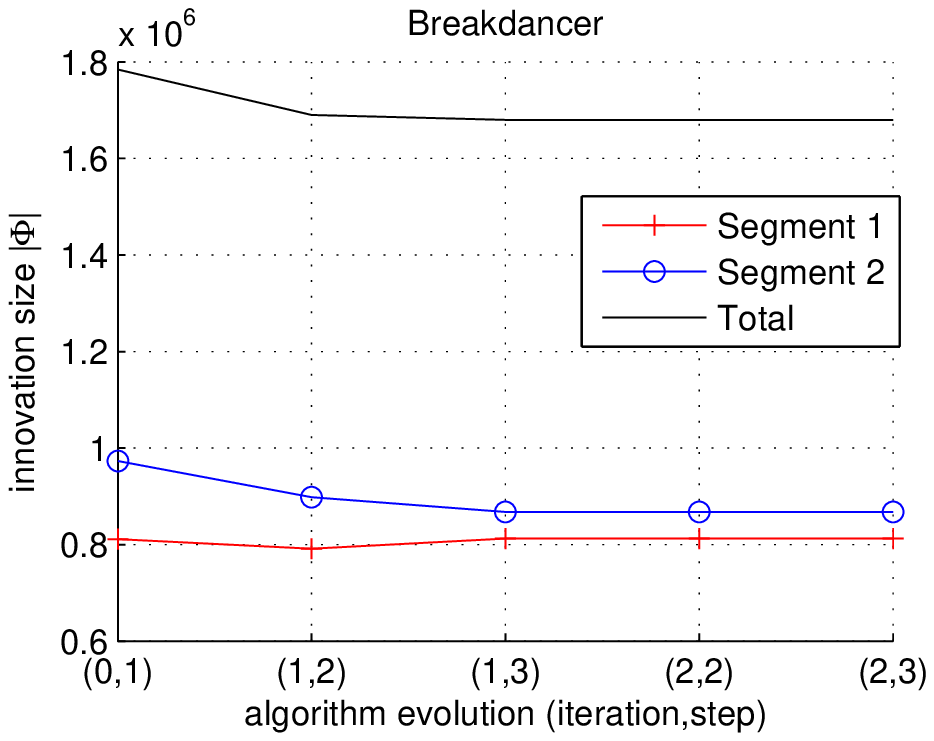,width=4cm}}
\end{minipage}

\vspace{1cm}

\begin{minipage}{0.48\linewidth}
 \centerline{\epsfig{figure= 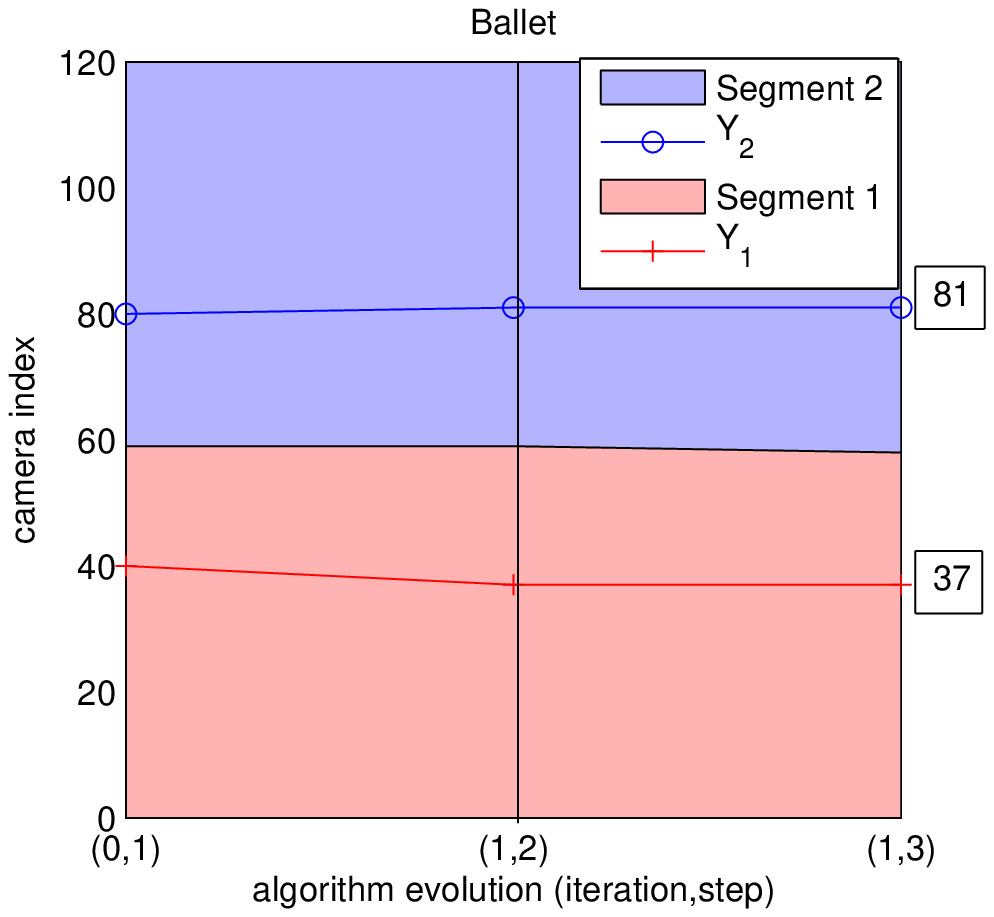,width=4cm}}
\end{minipage}
\begin{minipage}{0.48\linewidth}
 \centerline{\epsfig{figure= 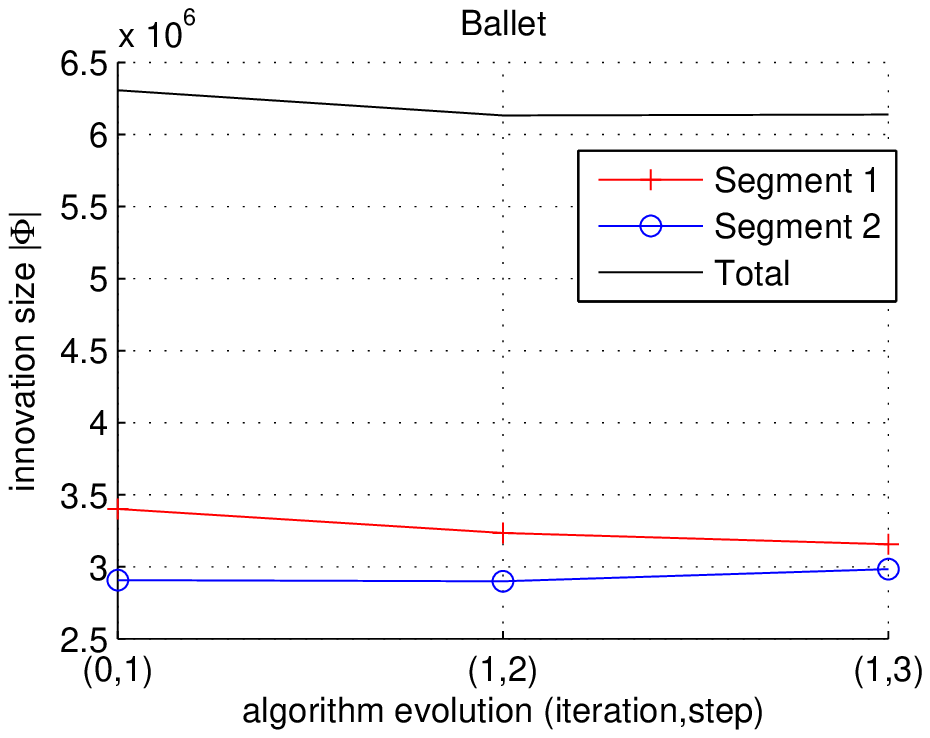,width=4cm}}
\end{minipage}
\begin{minipage}{0.48\linewidth}
\centerline{(a)}
\end{minipage}
\begin{minipage}{0.48\linewidth}
\centerline{(b)}
\end{minipage}
\caption{Partitioning results for two 1D  navigation segments when the initial reference frames are set at positions $40$ and $80$ (camera indexes). On the left, the evolution of the partitioning is illustrated as a function of the computation steps expressed as $(a,b)$, where $a$ is the iteration number and $b$ is the step. On the right, the evolution of the segment innovation size is illustrated.}
\label{fig:40_80}
\end{figure}

\begin{figure*}[htb]
  \centering
\begin{minipage}{0.48\linewidth}
 \centerline{\epsfig{figure= 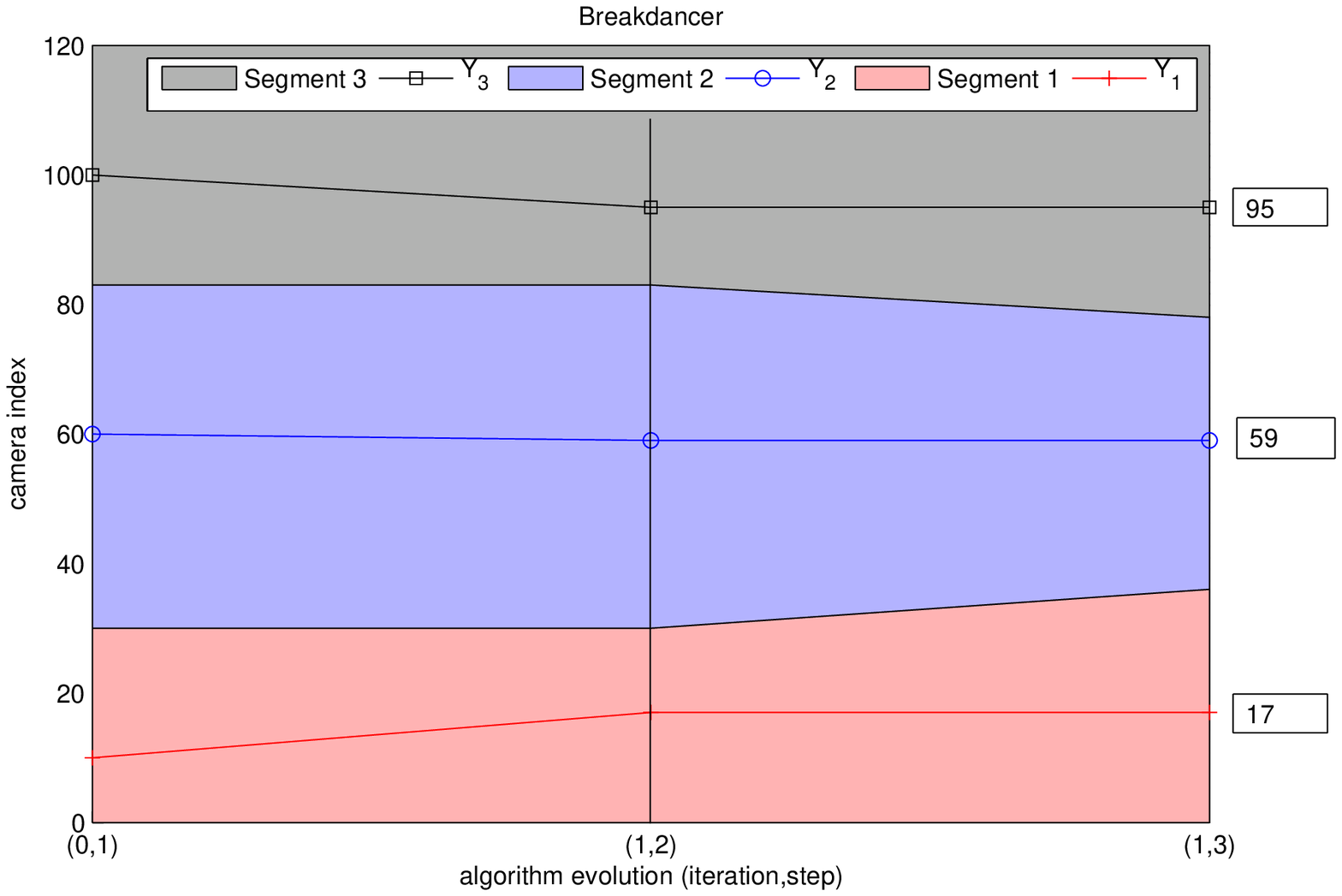,width=8cm}}
\end{minipage}
\begin{minipage}{0.48\linewidth}
 \centerline{\epsfig{figure= 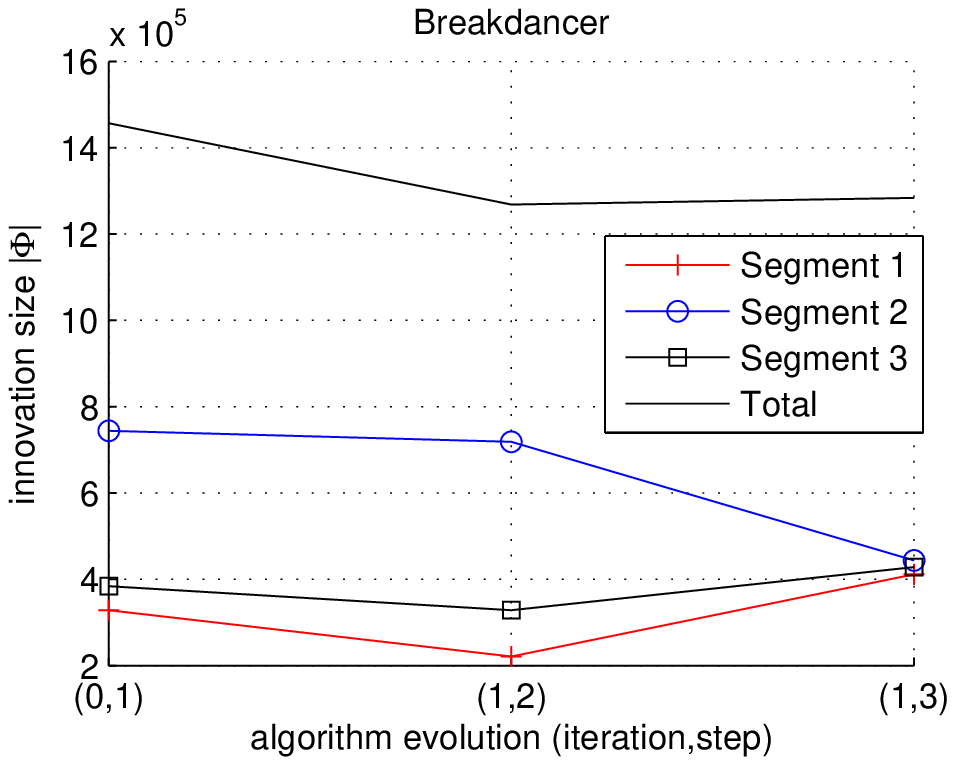,width=5cm}}
\end{minipage}

\vspace{1cm}

\begin{minipage}{0.48\linewidth}
 \centerline{\epsfig{figure= 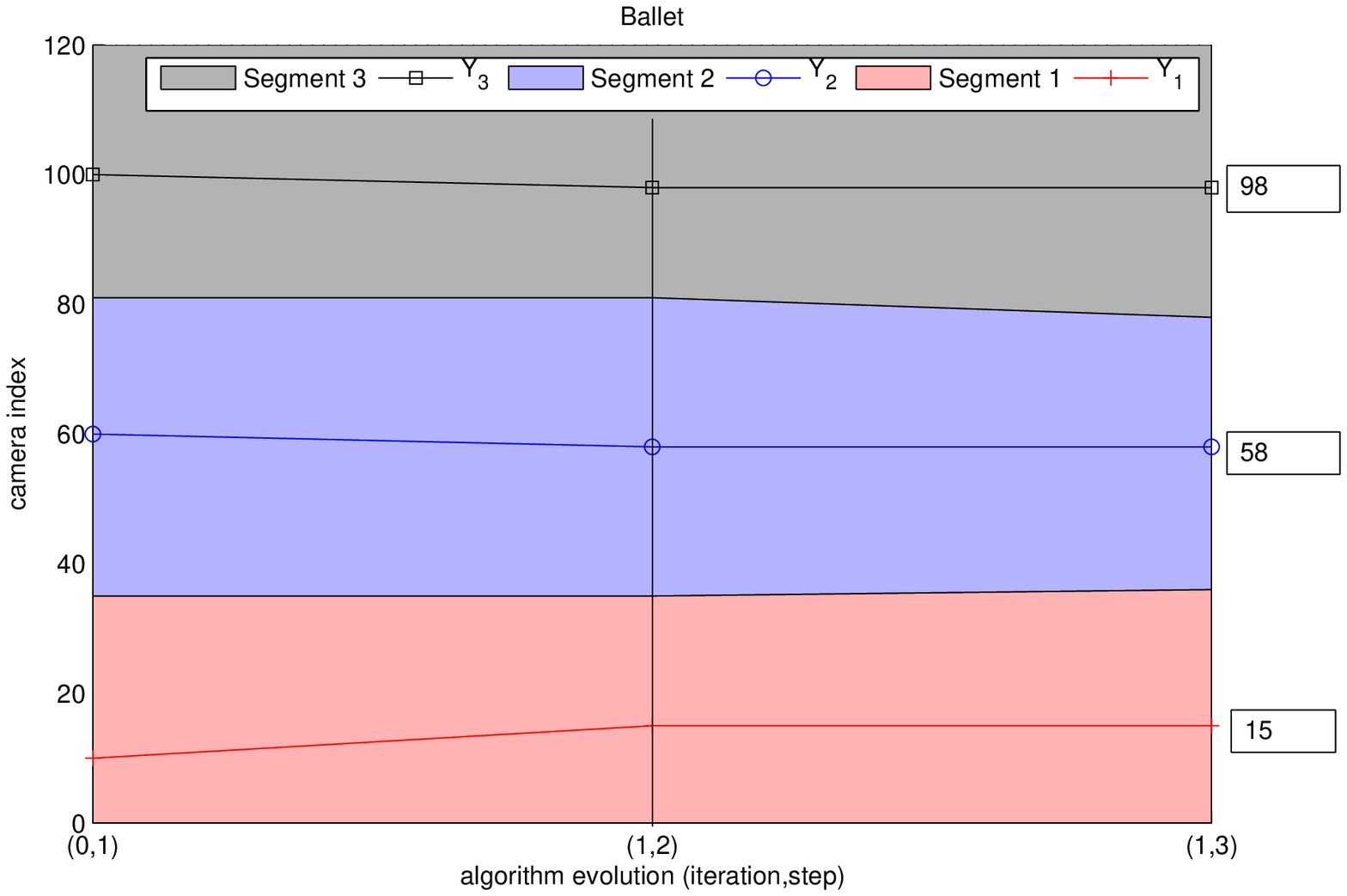,width=8cm}}
\end{minipage}
\begin{minipage}{0.48\linewidth}
 \centerline{\epsfig{figure= 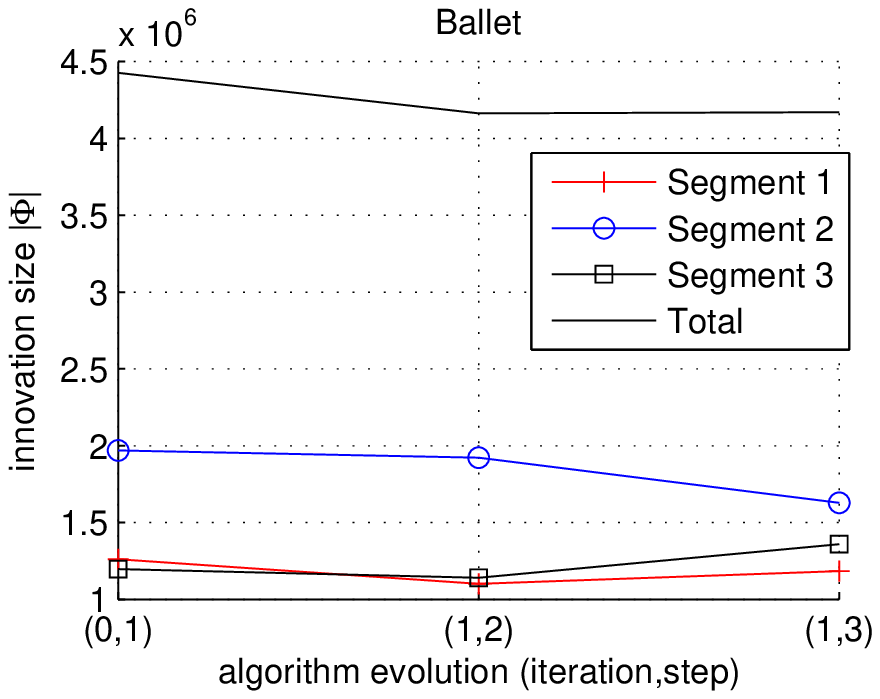,width=5cm}}
\end{minipage}
\begin{minipage}{0.48\linewidth}
\centerline{(a)}
\end{minipage}
\begin{minipage}{0.48\linewidth}
\centerline{(b)}
\end{minipage}
\caption{Partitioning results for three 1D navigation segments when the initial reference frames are set at positions $10$, $60$ and $100$ (camera indexes). On the left, the evolution of the partitioning is illustrated as a function of the computation steps expressed as $(a,b)$, where $a$ is the iteration number and $b$ is the step index. On the right, the evolution of the segment innovation size is illustrated.}
\label{fig:10_60_80}
\end{figure*}

\begin{figure*}[htb]
\begin{minipage}{0.6\linewidth}
 \centerline{\epsfig{figure= 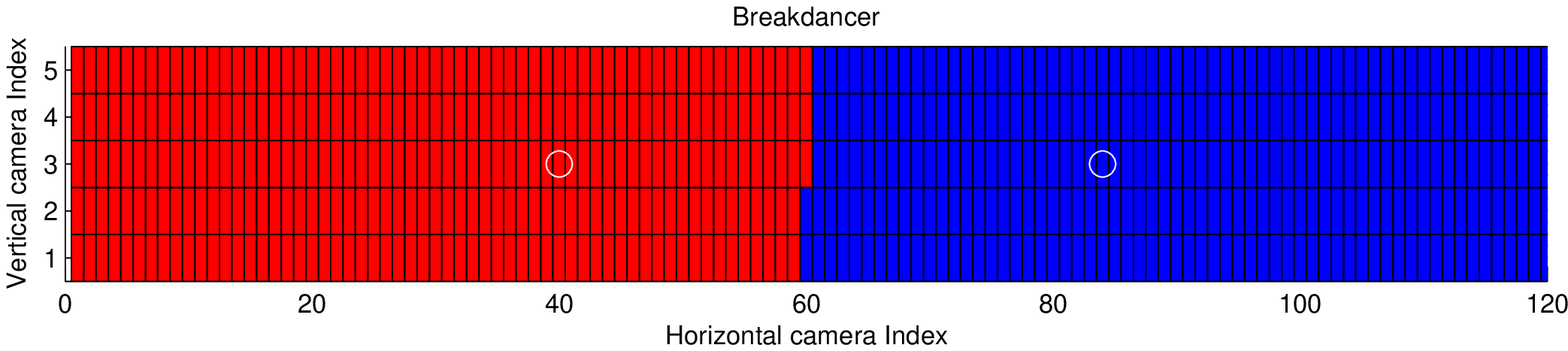,width=12cm}}
\centerline{(a)}
\end{minipage}
\begin{minipage}{0.48\linewidth}
 \centerline{\epsfig{figure= 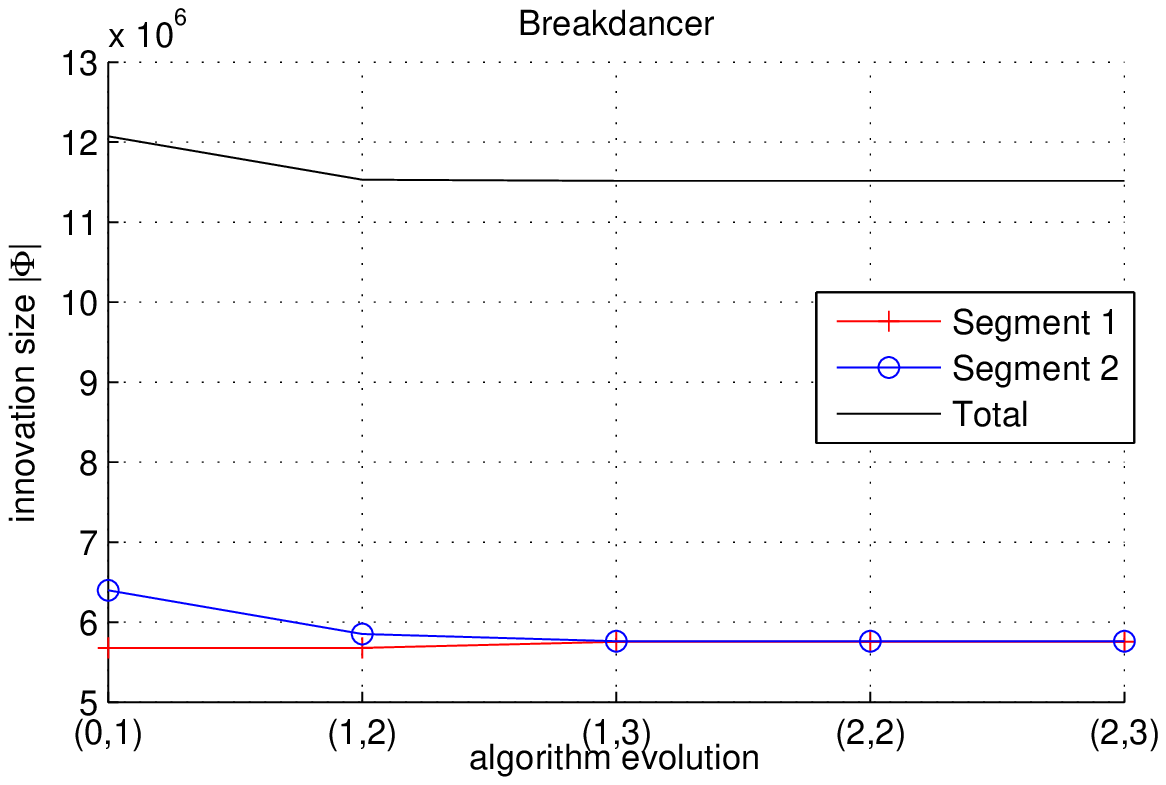,width=5cm}}
\centerline{(b)}
\end{minipage}
\caption{2D partitioning results for two navigation segments with initial reference frames at positions $(3,30)$ and $(3,60)$ (camera indexes). On the left, the final partitioning is illustrated; on the right, the  evolution of the innovation size $|\Phi|$ is shown as a function of the computation steps expressed as $(a,b)$, where $a$ is the iteration number and $b$ is the step index.}
\label{fig:30_60_2D}
\end{figure*}

\begin{figure*}[htb]
\begin{minipage}{0.6\linewidth}
 \centerline{\epsfig{figure= 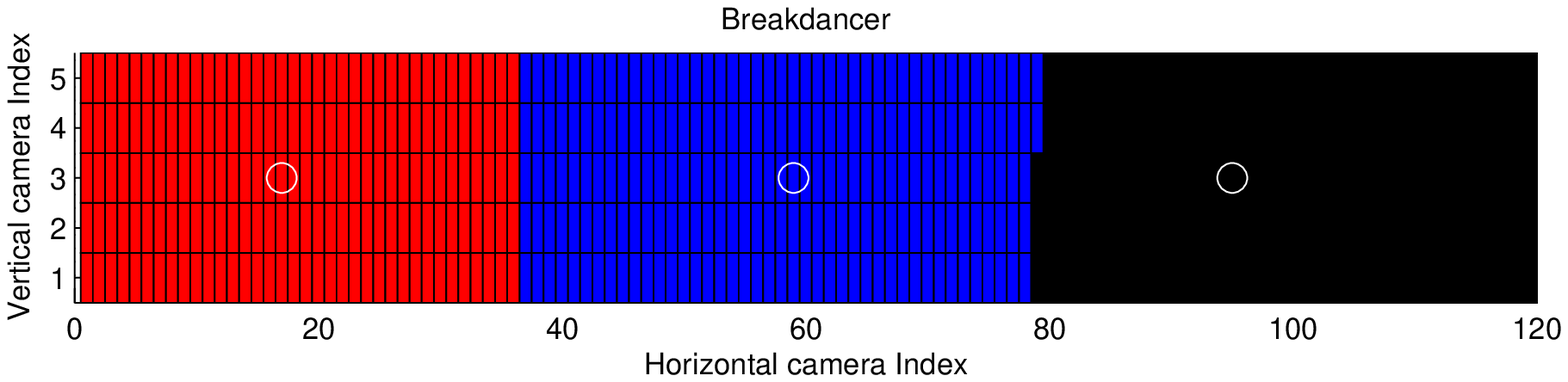,width=12cm}}
\centerline{(a)}
\end{minipage}
\begin{minipage}{0.48\linewidth}
 \centerline{\epsfig{figure= 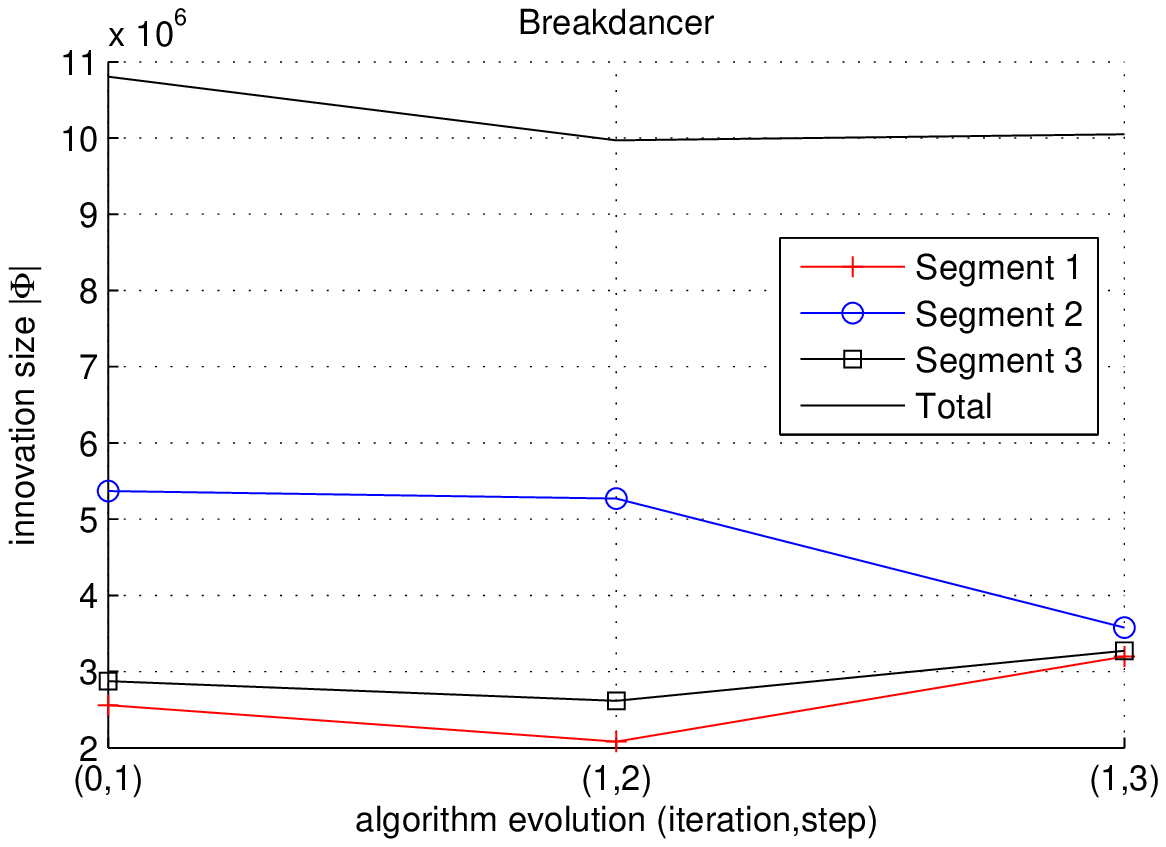,width=5cm}}
\centerline{(b)}
\end{minipage}
\caption{2D partitioning results for three navigation segments with initial reference frames at positions $(3,10)$, $(3,60)$ and $(3,100)$ (camera indexes). On the left, the final partitioning is illustrated; on the right, the  evolution of the innovation size $|\Phi|$ is shown as a function of the computation steps expressed as $(a,b)$, where $a$ is the iteration number and $b$ is the step index.}
\label{fig:10_60_100_2D}
\end{figure*}

We discuss now the results of the optimized partitioning algorithm and its effect on the size of the  segment innovation $\Phi$. We assume here that the number of partitions $N_V$ is predetermined.  Since the shape of the criterion function in our optimization problem is not completely convex, one needs to be careful in the initialization of the algorithm in order to avoid local minima. We put the initial reference frames at equidistant positions (in terms of geometrical similarity), which has been shown experimentally to be a good initial solution. 
In Fig.~\ref{fig:40_80} and \ref{fig:10_60_80}, we show the performance of our algorithm in the partitioning of a 1D navigation domain.  In each of these figures we show (a) the evolution of the partitioning and (b) the evolution of the segment innovation $|\Phi|$ through the successive steps of the partitioning algorithm. We see that, for $N_V=2$ and $N_V=3$, the total segment innovation decreases. Then, in each case, the size of the innovation $\Phi$ converges to a similar value. We also remark that the algorithm converges in a small number of steps towards a non equidistant distribution of the reference frames. 

We show in Fig.~\ref{fig:30_60_2D} and \ref{fig:10_60_100_2D} similar results for the partitioning of a 2D navigation domain.  We illustrate the final 2D partitioning and the evolution of the segment innovation size along the successive steps of the iterative optimization algorithm. We can see that the algorithm converges quickly and decreases the size of the segment innovation $\Phi$. 
More precisely, with both \emph{Ballet} and \emph{Breakdancer} test sequences, the algorithm never requires more than 3 iterations to converge.  It is interesting to notice that the resulting partitioning does not correspond to an equidistant  distribution of the reference frames (in terms of camera parameter distance).
Indeed, an equidistant distribution would have given reference frames position at indexes $(3,30)$ and $(3,90)$ (for 2 navigation segments) and $(3,20)$, $(3,60)$ and $(3,100)$ (for 3 navigation segments), whereas our partitioning method optimally positions them at indexes $(3,40)$ and $(3,84)$ (for 2 navigation segments) and $(3,17)$, $(3,59)$ and $(3,95)$ for 3 navigation segments. This is due to the fact that the proposed algorithm takes into account the scene content  in the definition of the navigation segments.

The convergence speed depends on the scene complexity, but generally stays pretty good. The main limitation of the algorithm is that  calculation of $|\Phi|$ which is relatively expensive and takes up to 90\% of the overall computation time. However, one can consider some scene learning, modeling or heuristics to improve the computation efficiency in dynamic scenes or realtime applications.
%

\subsection{Optimal number of navigation segments}\label{sec:navsegnum}

\begin{figure}[htb]
  \centering
\begin{minipage}{1\linewidth}
 \centerline{\epsfig{figure= 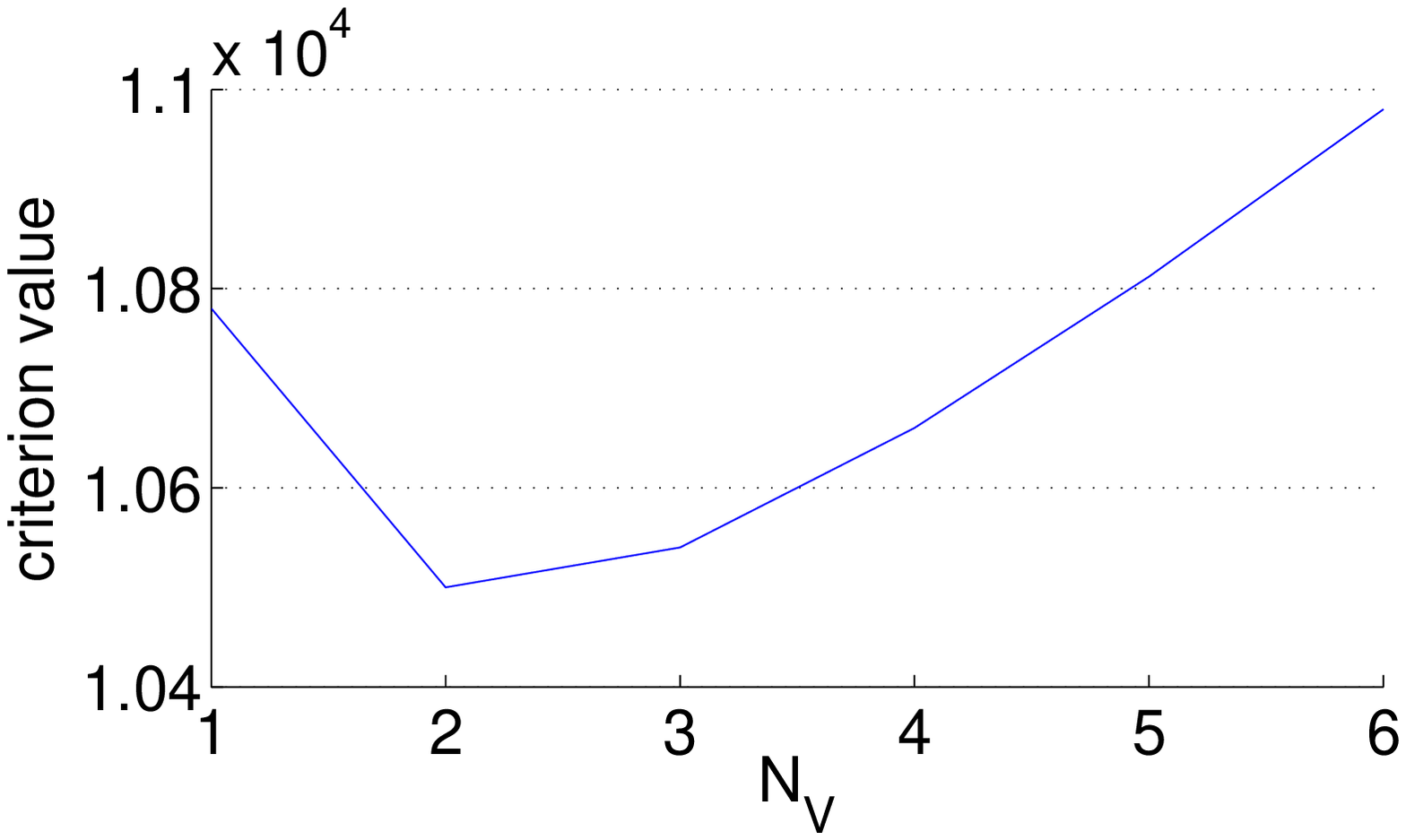,width=4cm}}
\centerline{(a) $\mu = 0.1$}
\end{minipage}
\begin{minipage}{1\linewidth}
 \centerline{\epsfig{figure= 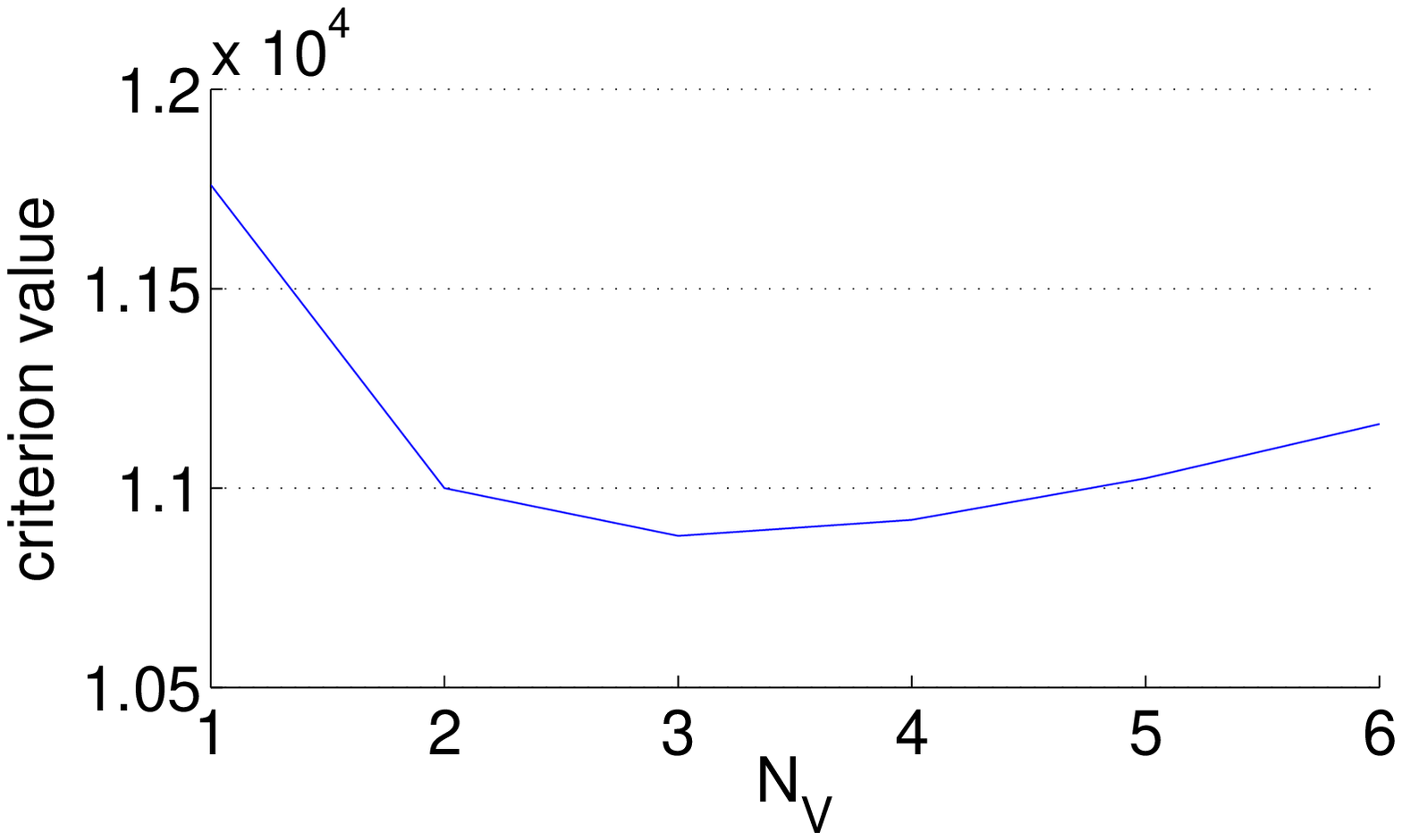,width=4cm}}
\centerline{(b) $\mu = 0.2$}
\end{minipage}
\begin{minipage}{1\linewidth}
 \centerline{\epsfig{figure= 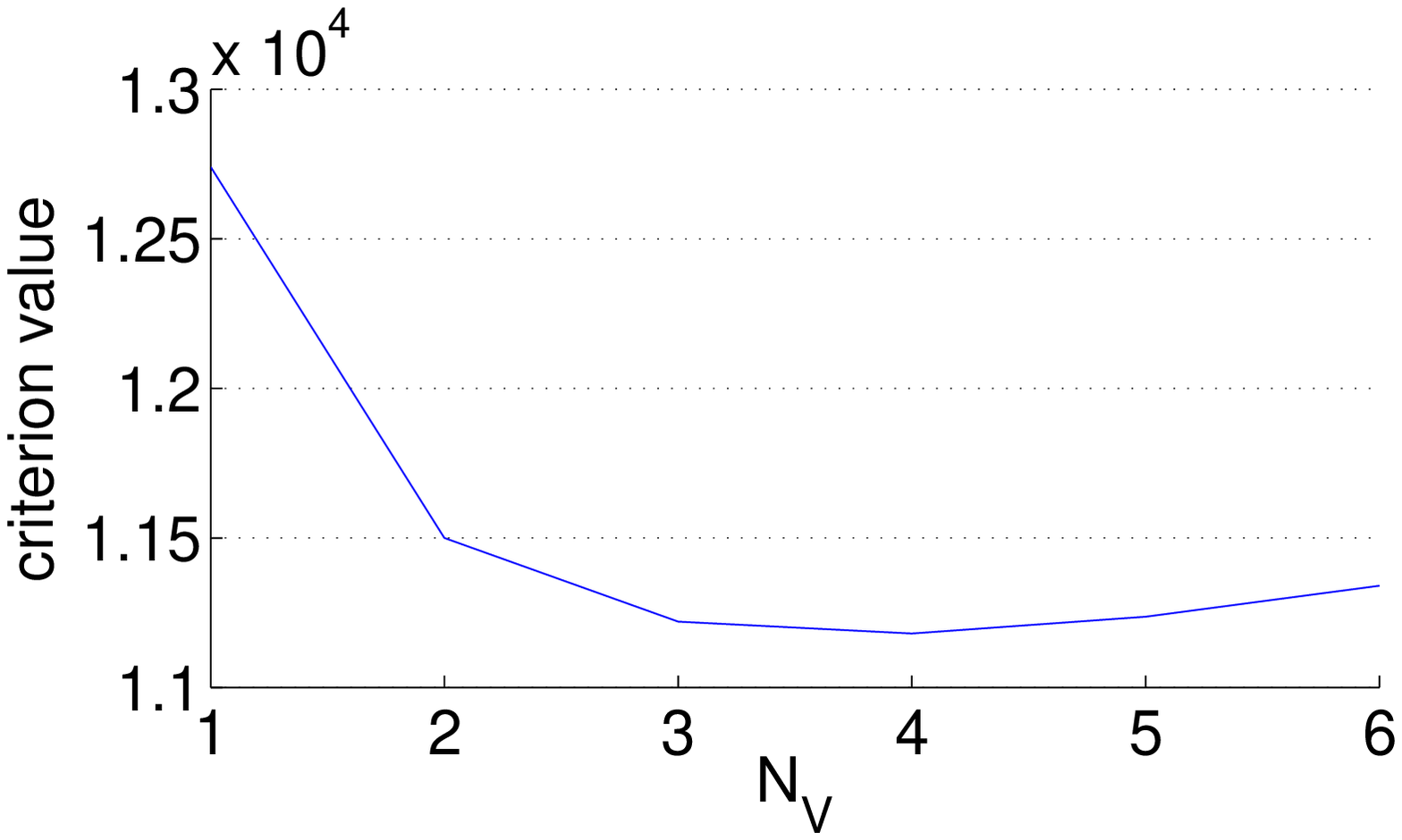,width=4cm}}
\centerline{(c) $\mu = 0.3$}
\end{minipage}
\caption{Optimal number of navigation segments $N_V^*$ for different values of relative weight-factor $\mu$ for \emph{Ballet} sequence.}
\label{fig:NV}
\end{figure}

We now study how the system determines the appropriate number of navigation segments. The optimal number of navigation segments $N_V^*$ is determined by minimizing the criterion given in Eq. (\ref{eq:NV}). We show in Fig.~\ref{fig:NV} the shape of this criterion function with different values of the relative weight factor $\mu$ in Eq.~(\ref{eq:numberOfPartitions}). In these tests, we have considered that the coding function $h$ is linear (\emph{i.e.,} $\varphi$ increases linearly with $\Phi$), as it is experimentally obtained in Fig.~\ref{fig:icipPhiandVarPhi}. Then, for each value of $N_V$, we estimate the storage and maximum rate costs for the \emph{Ballet} sequence. We see that we obtain different optimal number of navigation segments $N_V^*$ depending on the parameter $\mu$ that trades off storage and rate costs. We  further observe that, if $\mu$ is large (\emph{i.e.,} more importance is given to the rate cost), the algorithm selects a high number of navigation segments. On the contrary, if the storage cost has more importance, the system prefers a small number of navigation segments. The  parameter $\mu$ thus regulates the importance of the storage cost with respect to rate cost. This parameter is determined during the design of the system and depends on the system constraints and on the network delivery conditions.

\subsection{Rate and storage performance} \label{sec:rdeval}
\begin{figure}[htb]
  \centering
\begin{minipage}{0.6\linewidth}
 \centerline{\epsfig{figure= 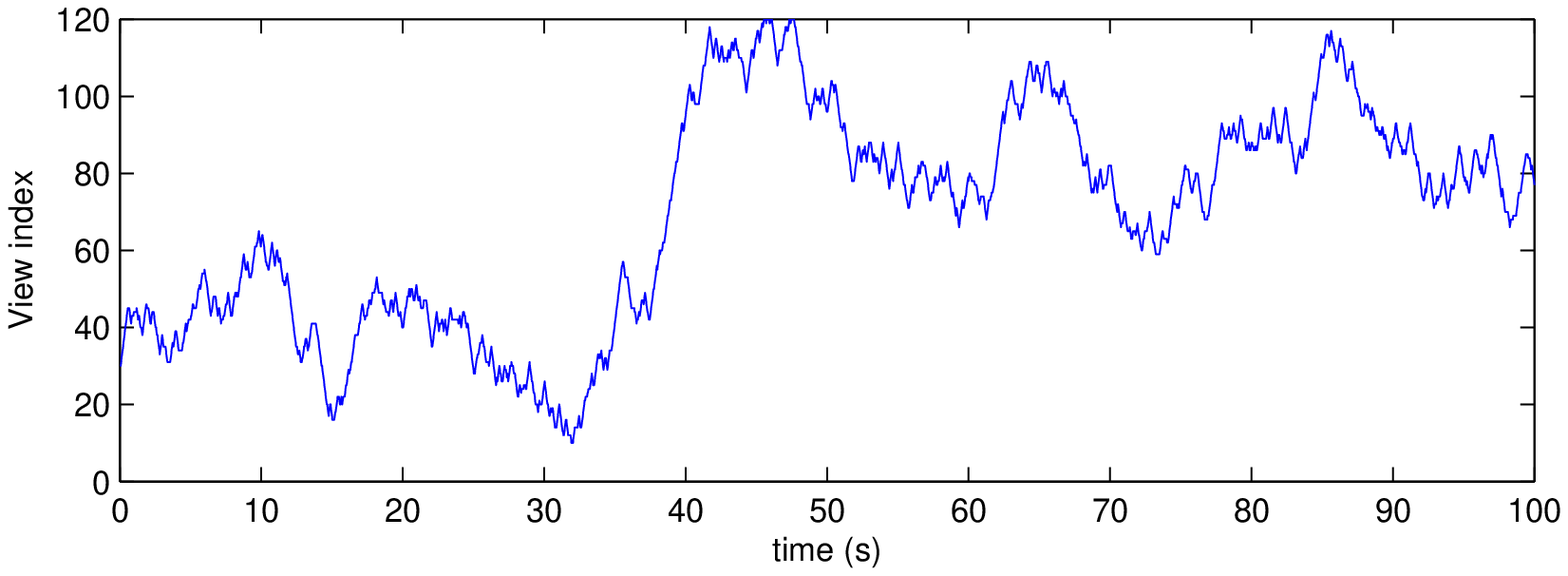,width=7cm}}
\centerline{(a) navigation path}
\end{minipage}
\begin{minipage}{0.6\linewidth}
 \centerline{\epsfig{figure= 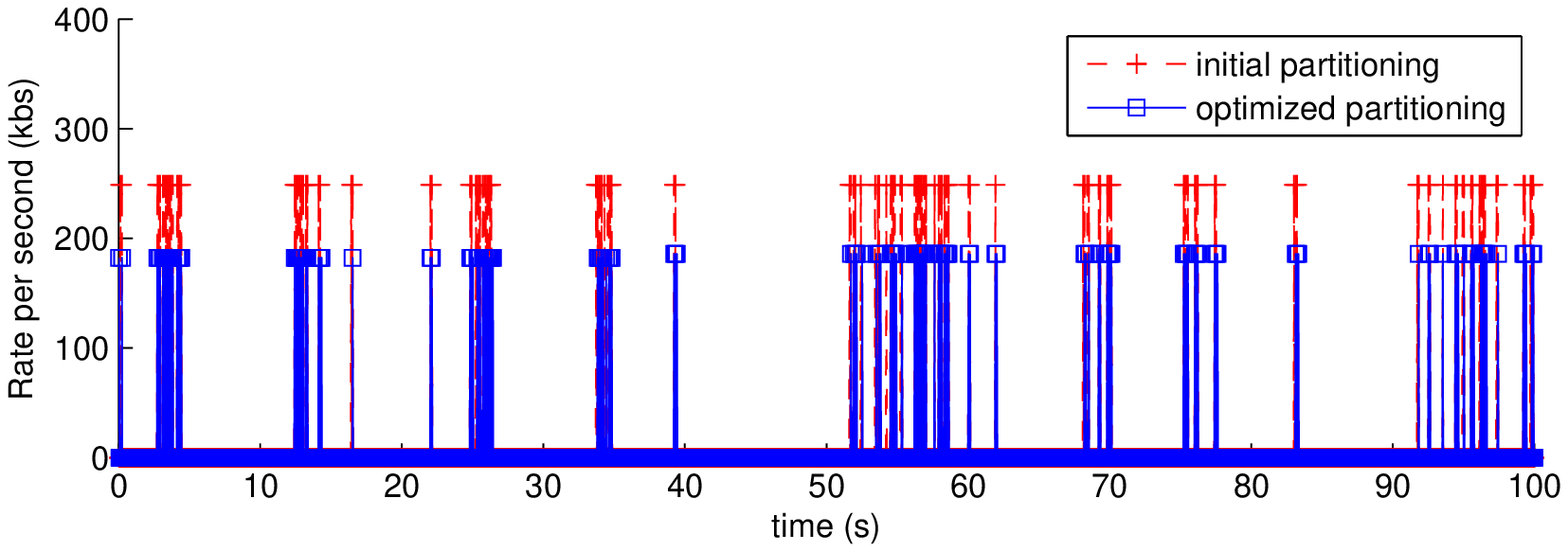,width=7cm}}
\centerline{(b) rate per second}
\end{minipage}
\begin{minipage}{0.6\linewidth}
 \centerline{\epsfig{figure= 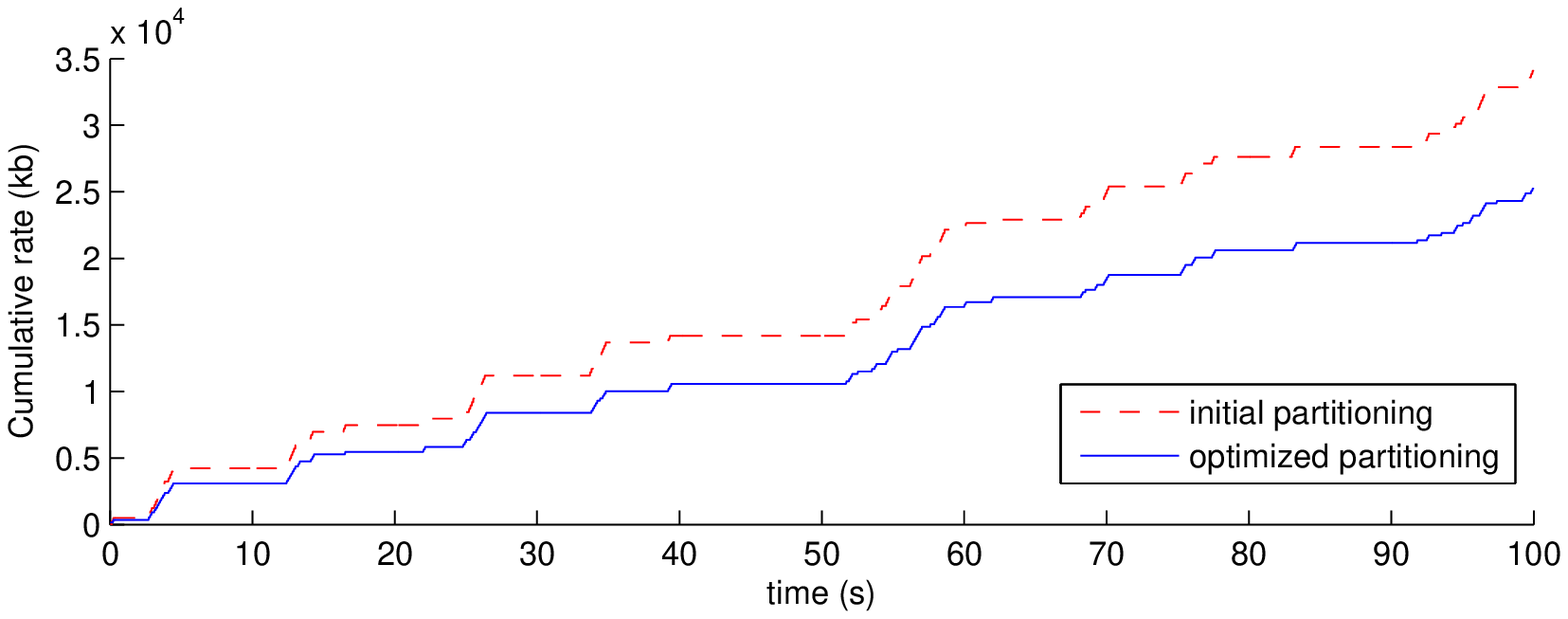,width=7cm}}
\centerline{(c) cumulative rate}
\end{minipage}
\caption{Rate cost performance with partitioning in $N_V = 3$ navigation segments of the sequence \emph{Ballet}, (with the partitioning solutions illustrated in Fig \ref{fig:10_60_80}).}
\label{fig:RD_ex1}
\end{figure}

\begin{figure}[htb]
  \centering
\begin{minipage}{1\linewidth}
 \centerline{\epsfig{figure= 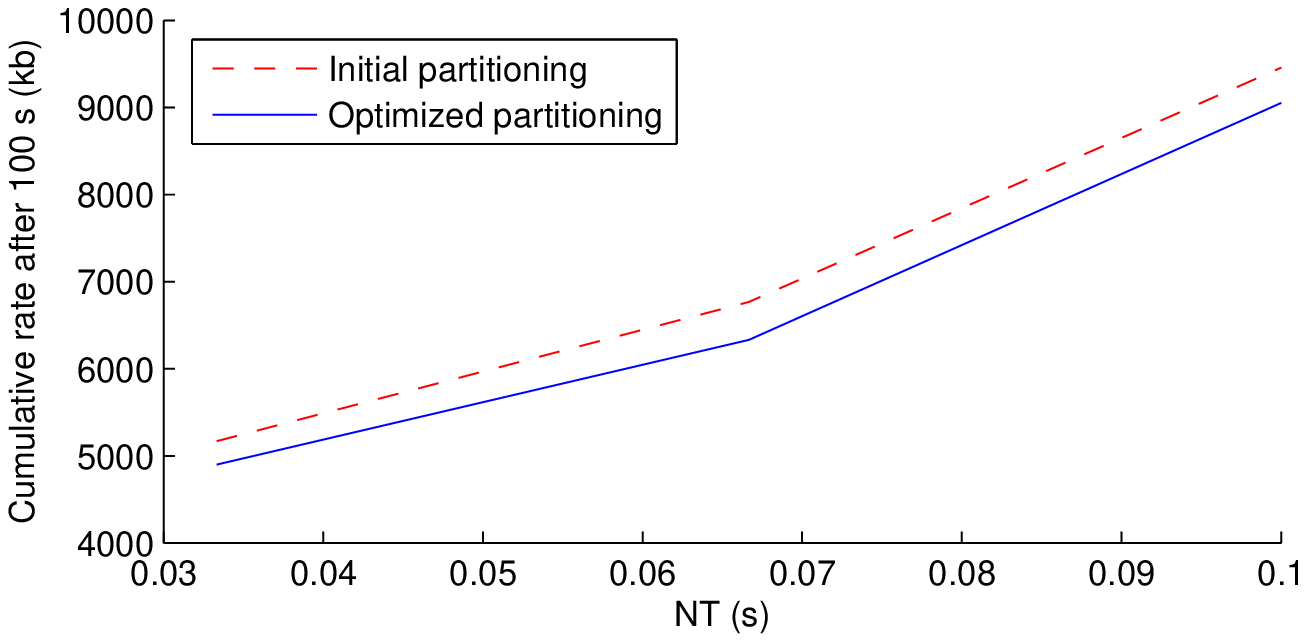,width=6cm}}
\centerline{(a) $N_V = 2$}
\end{minipage}
\begin{minipage}{1\linewidth}
 \centerline{\epsfig{figure= 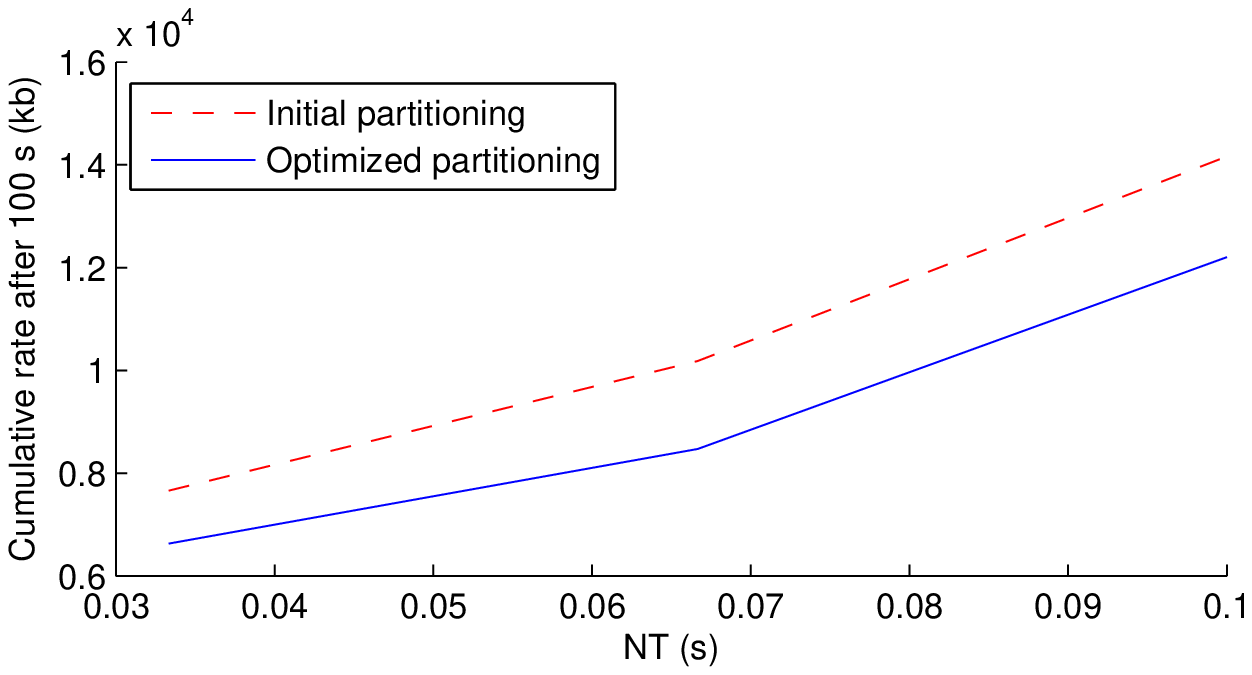,width=6cm}}
\centerline{(b) $N_V = 3$}
\end{minipage}
\caption{Averaged cumulative rate after 100s of navigation, for 100 navigation paths and for different values of $N_T$ (with the partitioning solutions for the \emph{Ballet} sequence illustrated in respectively Figs \ref{fig:40_80} and \ref{fig:10_60_80}).}
\label{fig:RD_avg}
\end{figure}

So far, we have mainly presented partitioning results in terms of the size of the segment innovation $|\Phi|$, which is directly related to the rate and storage costs. We now present  results that illustrate the performance of our algorithm in terms of rate values. We encode the auxiliary information with a quantized DCT representation as introduced in Sec.~\ref{sec:implementation}, which leads to a linear relation between the rate and the size $|\Phi|$ (as illustrated in Fig.~\ref{fig:icipPhiandVarPhi}). 
We  first model a possible navigation path for a user navigation of a duration of 100s. Each time, the path randomly stays on the same view (probability of $0.4$) or switches right or left (probability of $0.3$ each). The obtained path is represented in Fig.~\ref{fig:RD_ex1}(a). For this navigation path, we  simulate the communication of the client with the server and we  plot the evolution of the bit rate at the client in Fig.~\ref{fig:RD_ex1}(b), with the initial partitioning and the partitioning optimized with our algorithm. This bit rate per second is obtained by calculating the navigation segment sizes required during each second of user navigation. Here, the initial partitioning corresponds to the regular distribution of reference frames at the initialization of the optimization algorithm. We further plot the cumulative rate of the navigation process as a function of time for both partitioning solutions.
We see that the rate significantly decreases with the optimal partitioning; similarly the cumulative rate after $100$s of navigation is also smaller when the partitioning is optimal.  Similar results have been obtained for different navigation paths and different values of the number of navigation segments $N_V$.
To generalize these results, we have averaged the cumulative rate after 100s for $100$ navigation paths, and for different values of $N_T$ (time between two requests). We show the results in  Fig.~\ref{fig:RD_avg}.
We can see from all these representative results that the partitioning optimization leads to significant rate cost reductions. This validates our partitioning optimization solution.

\begin{figure}[htb]
  \centering
\begin{minipage}{0.48\linewidth}
 \centerline{\epsfig{figure= 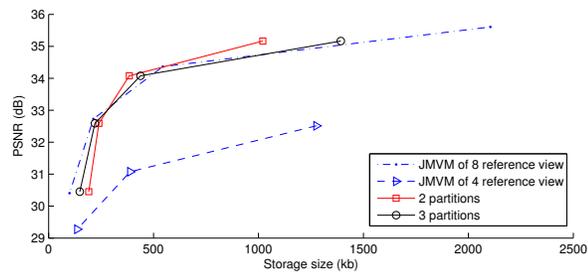,width=9cm}}
\end{minipage}
\caption{Distortion of a views $8,    23,    38,   53,    68,    83,    98,   113$ as a function of the storage size for two proposed partitioning solutions ($N_V = 2$ and $N_V=3$). It is compared to two solutions where the $8$ (and $4$) captured reference views are compressed jointly with JMVM (with no auxiliary information).}
\label{fig:JMVMComp}
\end{figure}

\begin{figure}[htb]
  \centering
 \centerline{\epsfig{figure= 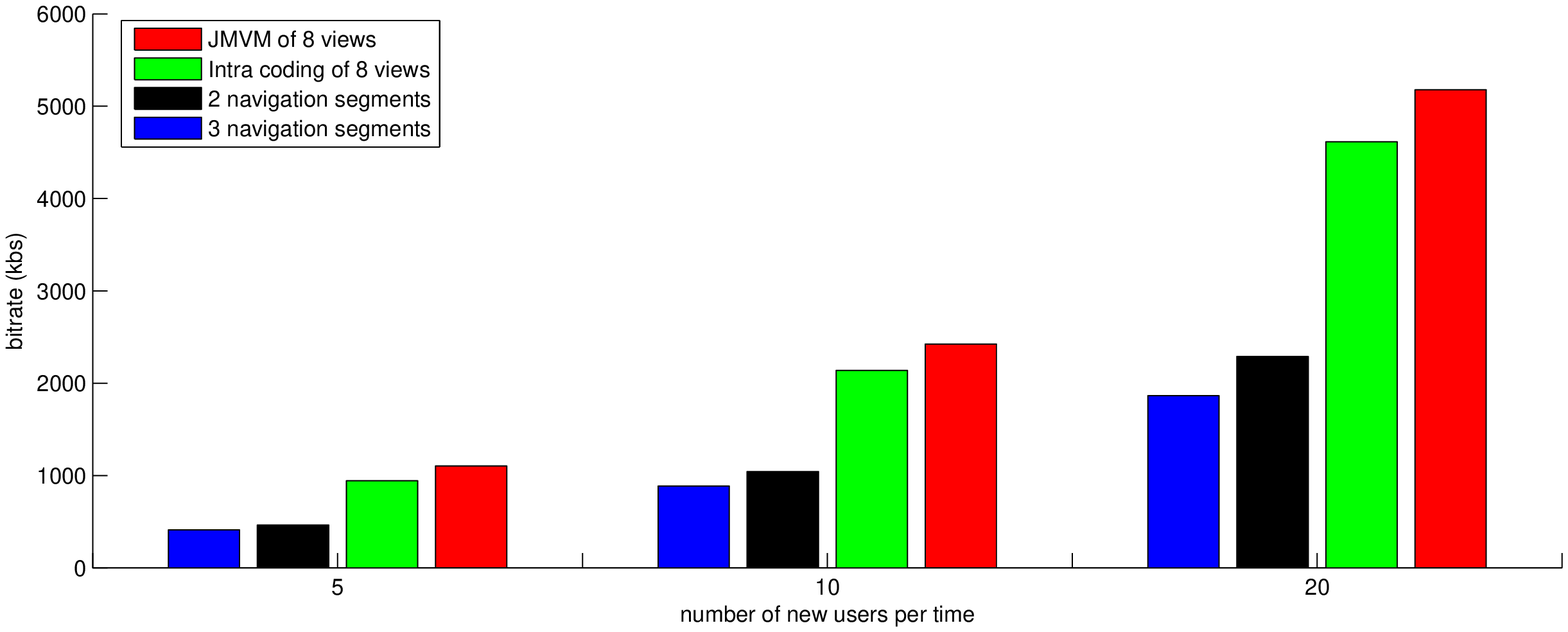,width=10cm}}
\caption{Simulation of a communication between one server and multiple users, during $1000$ seconds, for the \emph{Ballet} sequence. We compare different representation methods and plot the transmission rate (kbs) between the server and the users, for different numbers of new users $N_{nu}$.}
\label{fig:kbsVSnuser}
\end{figure}

\begin{figure}[htb]
  \centering
 \centerline{\epsfig{figure= 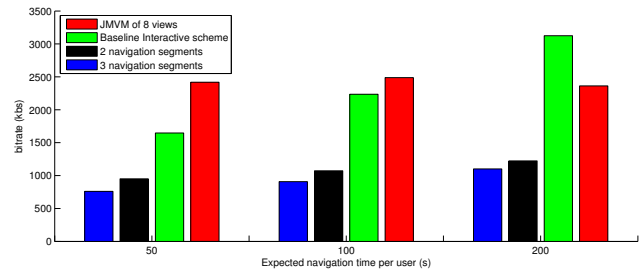,width=10cm}}
\caption{Simulation of a communication between one server and multiple users, during $1000$ seconds, for the \emph{Ballet} sequence. We compare different representation methods and plot the transmission rate (kbs) between the server and the users, for different times of navigation per user $T$.}
\label{fig:kbsVStol}
\end{figure}

Finally, in order to figure out the efficiency of the proposed representation method in terms of compression performance, we compare the storage cost of the proposed system with a baseline solution which is not adapted to interactivity. The latter consists in jointly compressing the camera views  with JMVM \cite{jmvm}, and in interpolating the other frames with bidirectional DIBR, as it is classically done in the literature. In our framework, we use two different partitioning solutions ($N_V=2$ and $N_V=3$) and we use the DCT-based auxiliary information coding explained in Sec.~\ref{sec:implementation}. The storage cost calculated contains the transmission of the reference image (color and depth) and the auxiliary information (for our solution). We compare both solutions and estimate the image quality of a representative sample set of images ($8,    23,    38,   53,    68,    83,    98,   113$ and the reference views). The results  are shown in Fig.~\ref{fig:JMVMComp} where we see that the proposed representation obtains similar compression performance as JMVM for $8$ camera views without auxiliary information. Such comparison is not conclusive about the potential for navigation since it provides storage costs only. In that sense, the experiment is  encouraging, as our framework reaches similar coding performance as a scheme that purely targets compression, while it also enables interactivity, which is not the case of JMVM.
We thus propose another comparison between our approach and some baseline methods. We compare our partition-based approach with two techniques. As in the previous test, we first consider the transmission of the whole set of reference frames jointly compressed with JMVM. Since this prediction scheme introduces strong inter-view dependencies, all views are all transmitted at the same time to enable view switching among all the frames of the navigation domain. Second, we propose to study an approach that encodes all of these frames independently with H.264/Intra. No auxiliary information is sent, and the decoder requires two reference frames to generate a virtual viewpoint. We have simulated a communication between one server and multiple users, during a certain time. At every second a given number of users, $N_{nu}$, arrive and start a navigation for a random duration with expected value $T$. This navigation is simply modeled with transition probabilities (at each instant, the probabilities of switching to the left or right views are set to $0.3$). We measure the total transmission rate between the server and the users for the three representation methods at similar image quality. We vary the number of new users per second  $N_{nu}$ in Fig.~\ref{fig:kbsVSnuser} and the expected time of navigation $T$ in Fig.~\ref{fig:kbsVStol}. We see that the optimal partitioning of the navigation domain  significantly reduces the bandwidth transmission rate compared to traditional image-based  data representation approaches. In other words, interactive schemes require new data representation because image-based description methods are not suited to random view transmission, as we have shown in Fig.~\ref{fig:kbsVSnuser} and \ref{fig:kbsVStol}.

\section{Conclusion}
In this paper, we propose a novel data representation method for interactive multiview imaging.  It is based on the notion of navigation domain, which is optimally split into several navigation segments. Each of these navigation segments is described with one reference image and auxiliary information, which enables a high quality user navigation at the receiver. In addition to this novel representation framework, we have proposed a solution for effective partitioning of the navigation domain and for selecting the best position for reference images. Experimental results show that the viewing experience of the user is significantly improved with a reasonable rate and storage cost. The comparison with common image-based representation methods is very encouraging and outline the potential of our framework for emerging interactive multiview systems, since our method enables navigation without large penalty on compression performance. Our future work will mainly focus on the extension of our navigation framework to dynamic scenes in order to enable efficient interactive multiview  video viewing.


\IEEEpeerreviewmaketitle

\ifCLASSOPTIONcaptionsoff
  \newpage
\fi

\bibliographystyle{IEEEtran}
\bibliography{/Users/thomasmaugey/Documents/EPFL/bibli/abbr,/Users/thomasmaugey/Documents/EPFL/bibli/epfl}

\begin{thebibliography}{10}
\providecommand{\url}[1]{#1}
\csname url@samestyle\endcsname
\providecommand{\newblock}{\relax}
\providecommand{\bibinfo}[2]{#2}
\providecommand{\BIBentrySTDinterwordspacing}{\spaceskip=0pt\relax}
\providecommand{\BIBentryALTinterwordstretchfactor}{4}
\providecommand{\BIBentryALTinterwordspacing}{\spaceskip=\fontdimen2\font plus
\BIBentryALTinterwordstretchfactor\fontdimen3\font minus
  \fontdimen4\font\relax}
\providecommand{\BIBforeignlanguage}[2]{{%
\expandafter\ifx\csname l@#1\endcsname\relax
\typeout{** WARNING: IEEEtran.bst: No hyphenation pattern has been}%
\typeout{** loaded for the language `#1'. Using the pattern for}%
\typeout{** the default language instead.}%
\else
\language=\csname l@#1\endcsname
\fi
#2}}
\providecommand{\BIBdecl}{\relax}
\BIBdecl

\bibitem{Muller_K_2011_pieee_tdv_rudm}
K.~M\"uller, P.~Merkle, and T.~Wiegand, ``3d video representation using depth
  maps,'' \emph{Proc. IEEE}, vol.~99, no.~4, pp. 643--656, Apr. 2011.

\bibitem{Smolic_A_2005_pieee_int_tdvrct}
A.~Smolic and P.~Kauff, ``Interactive 3{D} video representation and coding
  technologies,'' \emph{Proc. IEEE}, vol.~93, no.~1, pp. 98--110, Jan. 2005.

\bibitem{Benzie_P_2007_tcsvt_sur_tdtvtt}
P.~Benzie, J.~Watson, P.~Surman, I.~Rakkolainen, K.~Hopf, H.~Urey, V.~Sainov,
  and C.~von Kopylow, ``A survey of {3DTV} displays: Techniques and
  technologies,'' \emph{IEEE Trans. on Circ. and Syst. for Video Technology},
  vol.~17, no.~11, pp. 1647 -- 1658, Nov. 2007.

\bibitem{Alatan_A_2007_tcsvt_sce_rttdtvs}
A.~Alatan, Y.~Yemez, U.~G\"ud\"ukbay, X.~Zabulis, K.~M\"uller, C.~Erdem,
  C.~Weigel, and A.~Smolic, ``Scene representation technologies for 3dtv - a
  survey,'' \emph{IEEE Trans. on Circ. and Syst. for Video Technology},
  vol.~17, pp. 1587--1605, 2007.

\bibitem{Smolic_A_2007_tcsvt_cod_atdtvs}
A.~Smolic, K.~M\"uller, N.~Stefaniski, J.~Ostermann, A.~Gotchev, G.~Akar,
  G.~Triantafyllidis, and A.~Koz, ``Coding algorithms for 3dtv - a survey,''
  \emph{IEEE Trans. on Circ. and Syst. for Video Technology}, vol.~17, pp.
  1606--1621, 2007.

\bibitem{Vetro_A_2011_tb_tdt_cst}
A.~Vetro, A.~Tourapis, K.~M\"uller, and T.~Chen, ``3d-tv content storage and
  transmission,'' \emph{IEEE Trans. on Broadcasting}, vol.~57, pp. 384--394,
  2011.

\bibitem{Holliman_N_2011_tb_thr_ddraa}
N.~Holliman, N.~Dodgson, G.~Favalora, and L.~Pocket, ``Three-dimensional
  displays: a review and applications analysis,'' \emph{IEEE Trans. on
  Broadcasting}, vol.~57, no.~2, pp. 362--371, Jun. 2011.

\bibitem{Wiegand_T_2003_tcsvt_ove_hvcs}
T.~Wiegand, G.~Sullivan, G.~Bjontegaard, and A.~Luthra, ``Overview of the
  {H.264/AVC} video coding standard,'' \emph{IEEE Trans. on Circ. and Syst. for
  Video Technology}, vol.~13, no.~7, pp. 560--576, Jul. 2003.

\bibitem{Lou_JG_2005_real_timvvs}
J.~Lou, H.~Cai, and J.~Li, ``A real-time interactive multi-view video system,''
  in \emph{Proc. ACM Int. Conf. on Multimedia}, Singapore, 2005, pp. 161--170.

\bibitem{Maugey_T_2012_picip_con_vsimi}
T.~Maugey, P.~Frossard, and G.~Cheung, ``Consistent view synthesis in
  interactive multiview imaging.'' in \emph{Proc. IEEE Int. Conf. on Image
  Processing}, Orlando, Florida, US, Oct. 2012.

\bibitem{Ohm_jr_jct3v_wp3dsd}
J.~Ohm, D.~Rusanovskyy, A.~Vetro, and K.~Muller, ``{JCT3V-B}1006 work plan in
  3{D} standards development,'' 2012.

\bibitem{Suzuki_T_2013_jvt3v_eidmdt}
T.~Suzuki, M.~Hannuksela, Y.~Chen, S.~Hattori, and G.~Sullivan, ``{JCT3V-C1001
  MVC} extension for inclusion of depth maps draft text,'' 2013.

\bibitem{Rusanovskyy_D_2013_jct3v_tdavctm}
D.~Rusanovskyy, F.~Chen, Z.~Zhang, and T.~Suzuki, ``{JCT3V-C1003 3D-AVC} test
  model 5,'' 2013.

\bibitem{Karczewicz_M_2003_tcsvt_sp_sifdh}
M.~Karczewicz and R.~Kurceren, ``The {SP}- and {SI}-frames design for
  {H.264/AVC},'' \emph{IEEE Trans. on Circ. and Syst. for Video Technology},
  vol.~13, no.~7, pp. 637--644, Jul. 2003.

\bibitem{Chen_Y_2009_jadvsp_eme_mvcstdvs}
Y.~Chen, Y.~Wang, K.~Ugur, M.~Hannuksela, J.~Lainema, and M.~Gabbouj, ``The
  emerging {MVC} standards for 3d video services,'' \emph{EURASIP J. on Adv. in
  Sign. Proc.}, vol. 2009, pp. 1--13, 2009.

\bibitem{kurutepe_E_2007_tcsvt_cli_dssmvitdtv}
E.~Kurutepe, M.~Civanlar, and A.~Tekalp, ``Client-driven selective streaming of
  multiview video for interactive {3DTV},'' \emph{IEEE Trans. on Circ. and
  Syst. for Video Technology}, vol.~17, no.~11, pp. 1558--1565, Nov. 2007.

\bibitem{Tekalp_AM_2007_ieee-spm_tdt_oipesmv}
A.~Tekalp, E.~Kurutepe, and M.~Civanlar, ``{3DTV} over {IP}: end-to-end
  streaming of multiview videos,'' \emph{IEEE Signal Processing Magazine},
  no.~6, pp. 77--87, Nov. 2007.

\bibitem{Liu_Y_2010_jvci_rdo_ismvme}
Y.~Liu, Q.~Huang, S.~Ma, D.~Zhao, and W.~Gao, ``{RD}-optimized interactive
  streaming of multiview video with multiple encodings,'' \emph{Journal on
  Visual Commun. and Image Repr.}, vol.~21, no. 5-6, pp. 1--10, Jul. 2010.

\bibitem{Kimata_H_2004_ntt_fre_vvcumvc}
H.~Kimata, M.~Kitahara, K.~Kamikura, and Y.~Yashima, ``Free-viewpoint video
  communication using multi-view video coding,'' \emph{NTT Technical Review},
  vol.~2, no.~8, pp. 21--26, Aug. 2004.

\bibitem{Shimizu_S_2007_tcsvt_vie_smvcutdwdm}
S.~Shimizu, M.~Kitahara, H.~Kimata, K.~Kamikura, and Y.~Yashima, ``View
  scalable multiview video coding using 3-d warping with depth map,''
  \emph{IEEE Trans. on Circ. and Syst. for Video Technology}, vol.~17, no.~11,
  pp. 1485--1495, Nov. 2007.

\bibitem{Cheung_G_2009_ipvw_gen_rfsims}
G.~Cheung and N.~Ortega, A.and~Cheung, ``Generation of redundant frame
  structure for interactive multiview streaming,'' in \emph{Internat. Packet
  Video Workshop}, Seattlle, WA, USA, May 2009.

\bibitem{Cheung_G_2011_tip_int_ssmvurfs}
G.~Cheung, A.~Ortega, and N.~Cheung, ``Interactive streaming of stored
  multiview video using redundant frame structures,'' \emph{IEEE Trans. on
  Image Proc.}, vol.~3, no.~3, pp. 744--761, Mar. 2011.

\bibitem{Petrazzuoli_G_2011_picip_usi_dscdibriimva}
G.~Petrazzuoli, M.~Cagnazzo, F.~Dufaux, and B.~Pesquet-Popescu, ``Using
  distributed source coding and depth image based rendering to improve
  interactive multiview video access,'' in \emph{Proc. IEEE Int. Conf. on Image
  Processing}, Brussels, Belgium, 2011.

\bibitem{Xiu_X_2011_picip_fra_soimvsbnd}
X.~Xiu, G.~Cheung, and J.~Liang, ``Frame structure optimization for interactive
  multiview video streaming with bounded network delay,'' in \emph{Proc. IEEE
  Int. Conf. on Image Processing}, Brussels, Belgium, Sep. 2011.

\bibitem{Cheung_NM_2006_pvicip_vid_cfpobdsc}
N.~Cheung, H.~Wang, and A.~Ortega, ``Video compression with flexible playback
  order based on distributed source coding,'' in \emph{Proc. SPIE Visual
  Commun. and Image Processing}, San Jose, California, USA, Nov. 2006.

\bibitem{Xiu_X_2012_tmm_del_cimvfvs}
X.~Xiu, G.~Cheung, and J.~Liang, ``Delay-cognizant interactive streaming of
  multiview video with free viewpoint synthesis,'' \emph{IEEE Trans. on
  Multimedia}, vol.~14, pp. 1109--1126, 2012.

\bibitem{Tanimoto_M_2011_ieee-spm_fre_vtv}
M.~Tanimoto, M.~Tehrani, T.~Fujii, and T.~Yendo, ``Free-viewpoint {TV},''
  \emph{IEEE Signal Processing Magazine}, vol.~11, pp. 67--76, 2011.

\bibitem{Tanimoto_2012_ieee-spm_ftv_fvt}
M.~Tanimoto, ``Ftv: Free-viewpoint television,'' \emph{IEEE Signal Processing
  Magazine}, vol.~27, no.~6, pp. 555--–570, Jul. 2012.

\bibitem{CohenOr_D_2002_jgmip_fun_sv}
D.~Cohen-Or and A.~Kaufman, ``Fundamentals of surface voxelization,''
  \emph{Journal of Graphical Models and Image Processing}, vol.~57, pp.
  5453--461, 2002.

\bibitem{Tian_D_2009_pspie_vie_sttdv}
D.~Tian, P.~Lai, P.~Lopez, and C.~Gomila, ``View synthesis techniques for 3d
  video,'' \emph{Proc. of SPIE, the Int. Soc. for Optical Engineering}, vol.
  7443, 2009.

\bibitem{Muller_K_2008_jivp_vie_satdvs}
K.~M\"uller, A.~Smolic, K.~Dix, P.~Merkle, P.~Kauff, and T.~Wiegand, ``View
  synthesis for advanced 3d video systems,'' \emph{EURASIP J. on Image and
  Video Proc.}, vol. 2008, 2008.

\bibitem{website_VVS}
\BIBentryALTinterwordspacing
 [Online]. Available: \url{http://www.cse.unr.edu/~bebis/CS791E/}
\BIBentrySTDinterwordspacing

\bibitem{Criminisi_A_2004_tip_reg_forebii}
A.~Criminisi, P.~Perez, and K.~Toyama, ``Region filling and object removal by
  exemplar-based image inpainting,'' \emph{IEEE Trans. on Image Proc.},
  vol.~13, no.~9, pp. 1200--1212, 2004.

\bibitem{Daribo_I_2010_mmsp_dep_aiinvs}
I.~Daribo and B.~Pesquet-Popescu, ``Depth-aided image inpainting for novel view
  synthesis,'' in \emph{IEEE Int. Workshop on Multimedia Sig. Proc.}, Saint
  Malo, France, Oct. 2010.

\bibitem{Vetro_A_2011_pieee_ove_smvcehms}
A.~Vetro, T.~Wiegand, and G.~Sullivan, ``Overview of the stereo and multiview
  video coding extensions of the {H}.264/{MPEG}-4 avc standards,'' \emph{Proc.
  IEEE}, vol.~99, no.~4, pp. 626--642, Apr. 2011.

\bibitem{Fehn_C_2004_pspie-sipr_dep_ibrctnatdtv}
C.~Fehn, ``Depth-image-based rendering (dibr), compression and transmission for
  a new approach on 3d-tv,'' \emph{Proc. SPIE, Stereoscopic Image Process.
  Render.}, vol. 5291, pp. 93--104, 2004.

\bibitem{Shade_J_1998_psiggraph_lay_di}
J.~Shade, S.~Gortler, L.~He, and R.~Szeliski, ``Layered depth images,'' in
  \emph{Proc. Int. Conf. on Computer graphics and interactive techniques}, New
  York, NY, USA, Jul. 1998.

\bibitem{web_microsoft_ballet_break}
\BIBentryALTinterwordspacing
 [Online]. Available:
  \url{http://research.microsoft.com/en-us/um/people/sbkang/3dvideodownload/}
\BIBentrySTDinterwordspacing

\bibitem{Gersho_A_1992_vec_qsc}
A.~Gersho and R.~Gray, \emph{Vector quantization and signal compression},
  R.~Gallager, Ed.\hskip 1em plus 0.5em minus 0.4em\relax Kluwer academic
  publishers, 1992.

\bibitem{jmvm}
{ISO/IEC MPEG \& ITU-T VCEG}, ``Joint multiview video model ({JMVM}),''
  Marrakech, Morocco, Jan.13-19 2007.

\end{thebibliography}

\end{document}